\documentclass[journal]{IEEEtran}
\usepackage{nicematrix}
\usepackage{color}
\usepackage{cite}
\usepackage{textcomp}
\usepackage{xcolor}
\usepackage{balance}
\usepackage{graphicx}
\usepackage[english]{babel}
\usepackage{amsthm,amssymb}
\usepackage{amsmath}
\usepackage{subcaption}
\usepackage{mathtools}
\usepackage{tikz}
\usepackage{pgf}
\usepackage{amsfonts}
\usepackage{bm}
\usepackage[bottom]{footmisc}
\usepackage[ruled,linesnumbered]{algorithm2e}
\usepackage[T1]{fontenc}
\usepackage{bbm}
\usepackage{enumerate}

\usepackage{setspace}
\usepackage{array}
\usepackage{multirow}
\usepackage{diagbox}

\DeclareMathOperator*{\argmax}{arg\,max}
\newtheorem{rem}{Remark}

\allowdisplaybreaks

\def\BibTeX{{\rm B\kern-.05em{\sc i\kern-.025em b}\kern-.08em
    T\kern-.1667em\lower.7ex\hbox{E}\kern-.125emX}}
\begin{document}

\title{Priority-aware grouping-based multihop routing scheme for RIS-assisted wireless networks}

\author{Authors}
\author{Lakshmikanta Sau, Priyadarshi Mukherjee,  \textit{Senior Member, IEEE,} and Sasthi~C.~Ghosh
\thanks{L. Sau, P. Mukherjee, and S. C. Ghosh are with the Advanced Computing \& Microelectronics Unit,  Indian Statistical Institute, Kolkata 700108, India. (E-mail: lakshmikanta\_r@isical.ac.in, priyadarshi@ieee.org, sasthi@isical.ac.in).
}}

\maketitle
\begin{abstract}
Reconfigurable intelligent surfaces (RISs) is a novel communication technology that has been recently presented as a potential candidate for beyond fifth-generation wireless communication networks. In this paper, we propose a priority-aware user traffic-dependent grouping-based multihop routing scheme for a RIS-assisted millimeter wave (mmWave) device-to-device (D2D) communication network with spatially correlated channels. Specifically, the proposed scheme exploits the priority of the users (based on their respective delay-constrained applications) and the aspect of spatial correlation in the narrowly spaced reflecting elements of the RISs. Here, based on the other users in the neighborhood, their respective traffic characteristics, and the already deployed RISs in the surroundings, we establish a multihop connection for information transfer from one of the users to its intended receiver. In this context, we take into account the impact of considering practical discrete phase shifts at the RIS patches instead of its ideal continuous counterpart. Moreover, we also claim and demonstrate that the existing classic least remaining distance (LRD)-based approach is not always the optimal solution. Finally, numerical results demonstrate the advantages of the proposed strategy and that it significantly outperforms  the existing benchmark schemes in terms of system performance metrics such as data throughput, energy consumption, as well as energy efficiency.

\end{abstract}
\section{Introduction}
In recent years, due to applications such as the Internet of Things (IoT) and enhanced mobile broadband communication, wireless traffic has increased exponentially; it is projected to expand by more than five times in the $2023-2028$ period \cite{ericsson}. As a result, various technologies such as adaptive modulation, beamforming, and multiple access frameworks have been developed in the last few decades to support this overwhelming data traffic. The unifying theme behind all these advancements is to intelligently exploit and adapt to the randomly fluctuating wireless channel instead of having a control over the same. On the contrary, a new technology termed as the reconfigurable intelligent surfaces (RISs) is aimed at controlling the random nature of the channel for the purpose of its maximum utilization \cite{risi2}. RISs essentially consist of arrays of reconfigurable passive elements embedded on a flat metasurface, which effectively `controls' the channel instead of adapting to its varying nature \cite{risrev}. These passive elements can be switched ON and OFF by adjusting the bias voltage. Moreover, without the need of any radio-frequency chains, they can reflect the incident signal in a desired direction. This results in reducing the implementation cost, which also enhances the system energy efficiency \cite{impl}. Furthermore, multiple variations of the RISs \cite{rev3_1,rev3_2,rev3_3} have also opened new directions of research. By incorporating imperfect channel state information and hardware impairments, the authors in \cite{rev3_1} investigates simultaneosuly transmitting and reflecting RIS-assisted non-orthogonal multiple access schemes. In \cite{rev3_2}, beamforming algorithms are proposed for enhancing the performance of reconfigurable holographic surfaces assisted  cell-free networks. The work in \cite{rev3_3} investigates active RIS-aided systems and analyzes its performance.

In this context, RIS-assisted device-to-device (D2D) communication \cite{risi3, new1,new2} forms a very relevant and interesting direction of research. The authors in \cite{risi3} focus on uplink D2D-enabled cellular networks assisted by an RIS, where multiple D2D links share the same spectrum. In \cite{new1}, the outage and rate performance of an RIS-assisted D2D communication underlaying cellular network is investigated by considering a generalized Nakagami-$m$ fading scenario. By using the concept of Markov service processes, the work in \cite{new2} analyzes the throughput of a delay-constrained RIS-assisted D2D communication scenario. Apart from system energy efficiency improvement by using RISs, the aspect of obtaining a high data rate can also be addressed by the usage of high frequency signals, for example, the millimeter waves (mmWaves) \cite{mm10}. Note that, although mmWaves solve the problem of enhanced data rate, it has its own set of shortcomings such as significant propagation and high penetration losses. As a result, RIS assisted mmWave-based D2D network appears to be the solution for such scenarios, especially where the direct line of sight (LoS) link is of inadequate strength. To fully exploit the advantages of such networks, the aspect of strategic RIS placement is extremely crucial as they need to have clear LoS links with both the devices that intend to communicate with each other. Optimal RIS placement \cite{sauris,deb2021ris,opris} can significantly extend the coverage of such networks with a minimum number of RISs, which effectively reduces both the hardware and installation expenditure.

We further note that the RIS is essentially a passive device that simply reflects the incoming signal as well as the noise to its desired direction by tuning its parameter. Moreover, any given pair of users always communicate with each other for only a finite amount of time. Thus, augmenting the environment with a large number of RISs may result in unnecessary wastage of resources. A possible solution to this problem is to adopt a cooperative multihop approach \cite{mhop1,mrirs}, i.e., apart from the RISs, the other intermediate users (IUs) present in the surroundings, depending on whether they are busy or idle, can be used as relays, namely, decode and forward (DF) or amplify and forward (AF). While AF relays have low processing cost, DF relays have high reliability as they forward only the decoded information portion to the next hop and not the entire received information-plus-noise mixture \cite{dfaf10}. Thus, the option of having the idle IUs to act as DF relays seems beneficial. However, the role of RISs cannot be totally overlooked as in case of delay-constrained scenarios and/or idle IU unavailability, they can provide the aspect of guaranteed reliability \cite{dramp}. Note that, the aspect of delay-constrained communication is very crucial, especially in 5G applications like ultra reliable low latency communication (URLLC) \cite{urllc}. In this context, \cite{mpsensor} proposes an energy-efficient adaptive topology-management system for sustaining network connectivity in resource-constrained wireless sensor networks (WSNs). Also, \cite{mprouting} investigates a fuzzy logic-based framework for quality-of-service aware routing protocol in wireless body area networks.

Accordingly, there exists some works, which investigate various aspects of routing in RIS-aided multihop networks. The work in \cite{DRL} proposes a practical hybrid beamforming architecture for multi-hop multiuser RIS-empowered wireless THz communication networks. Energy-efficiency (EE) being one of the critical issues nowadays, \cite{eecomp} investigates the aspect of EE maximization of the RIS-assisted multihop networks.

Moreover, as far as the structure of a RIS is concerned, certain aspects are critical to be considered and analyzed in this multihop framework. Firstly, the arrays of the reconfigurable elements are very narrowly spaced, which implies that the channels between the consecutive reflecting elements and the source/destination are spatially correlated \cite{corr,secure}. Hence, to avoid the huge channel estimation overhead in RIS-based systems, the works in \cite{partition3,grouping} propose grouping strategies where the nearby RIS elements are grouped together into smaller non-overlapping surfaces with identical phase shifts. Secondly, most existing works assume a continuous nature of the phase shifts that are offered by a RIS whereas in practical systems, they are actually discrete in nature \cite{how_much_phase}. Hence, these limited phase shifts do have a significant impact on the system performance. These aspects of a RIS cannot be overlooked if we are interested in investigating the practically feasible of a joint user and RIS based multihop framework. Furthermore, least remaining distance (LRD)-based approach is a standard procedure in literature to establish a multihop connection between two users, whether the sole objective is to reduce the effective distance between the two in each hop \cite{lrd2020}. But, unlike in this case, this approach stands valid if and only if we do not take the channel conditions and the availability of the IUs into consideration.

Motivated by these, in this paper, we consider the aspect of multihop routing in a grouping-based RIS-assisted wireless network with spatially correlated channels. Effective grouping implies that a single RIS can cater to multiple pairs of devices simultaneously. Between a given pair of devices, our aim is to find a sequence of appropriate RISs and IUs acting as relays when a direct LoS does not exist between them. Moreover, it may so happen that multiple relay requests arrive at a particular IU. In that case, the IU has to decide which request to serve in that instance. In this context, based on the respective delay constraint and the channel condition, we propose a metric that takes this decision. It is important to note that our priority is to make more use of the IUs over the RISs as explained earlier for delay-sensitive scenarios. Accordingly, we developed an algorithm to select the sequence of appropriate IUs and RISs to deliver the signals from source to destination. Finally, the numerical results demonstrate the benefits of the proposed framework, which results in enhanced data rate, reduced energy consumption, and enhanced energy efficiency, respectively, with respect to the existing benchmark schemes.
In light of this, the innovative contributions of this paper are summarized as follows:
\begin{itemize}
    \item By taking into account the aspect of spatial correlation, we characterize the achievable rate of grouping-based RIS systems. Accordingly, depending on the data rate requests of the users, a group selection criteria is presented.
    \item We propose an intelligent scheduling strategy to handle multiple requests arriving at an IU, by taking into account their delay constraints, respective channel conditions, and also the activity status (idle/busy) of the concerned IU.
    \item Instead of choosing the least remaining distance (LRD) as the criteria for selecting the next hop, we choose that particular IU as next hop, which provides the best data rate depending on the channel condition. In this context, we have used adaptive modulation for calculating the data rate obtained among the IUs.
    \item Finally, to characterize the performance of our proposed approach, we offer both theoretical and simulation results to demonstrate that ours outperforms the conventional LRD-based as well as the existing benchmark schemes.
\end{itemize}

\begin{figure}[t]
    \centering
    \includegraphics[width=\linewidth]{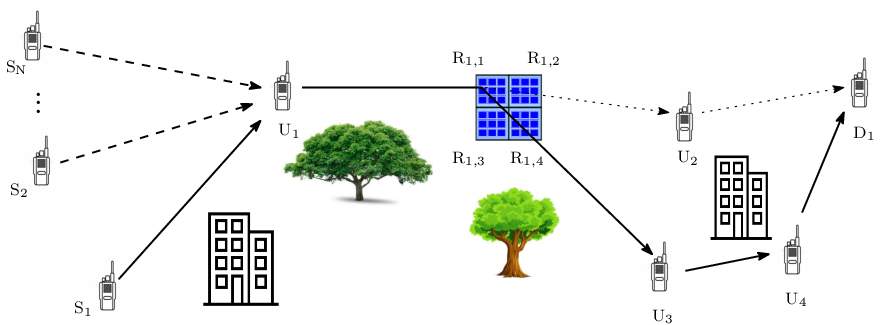}
    \vspace{-2mm}
    \caption{\footnotesize System Model}
    \label{sys}
    \vspace{-2mm}
\end{figure}

The rest of this paper is organized as follows. Section \ref{smod} describes the system model, Section \ref{prop} presents the proposed strategy, section \ref{ufwork} describes the analysis of the proposed scheduling strategy, the numerical results are presented in Section \ref{numerical} and finally, Section \ref{con} concludes the work.

\section{System Model}\label{smod}
\subsection{Network Topology}  \label{nt}
The considered topology consists of $N$ sources $S_1,\cdots,S_N$ and their respective destinations $D_1,\cdots,D_N$, $K$ RISs $R_1,\cdots,R_K$, and $M$ IUs $U_1,\cdots, U_M$. In a delay-constrained scenario, all the sources transmit signals with constant power $P$,  where the maximum acceptable delay for $S_n-D_n$ pair is $T_{d_n}$ $\forall$ $n=1,\cdots, N$. The sources are prioritized based on their acceptable delay limit, i.e., higher $T_{d_n}$ implies lower priority, and vice-versa. We consider all such $S_n-D_n$ pair for which the direct LoS link does not exist. We further assume that the devices cannot communicate beyond a distance $r$, each IU has a sufficient capacity buffer, and all D2D pairs adhere to time-slotted synchronous communication \cite{sync} with slot duration $T_s$. Table \ref{tabrsta} shows the list of variables used, along with the descriptions.

\begin{table*}[!t]
\centering
 \caption{\footnotesize SUMMARY OF NOTATIONS.} \label{tabrsta}
\resizebox{2\columnwidth}{!}{%
  \begin{tabular}{|c|c||c|c|}
    \hline \hline
    \textbf{Notation} & \textbf{Description} &  \textbf{Notation} &  \textbf{Description}\\
    \hline
     $K,M$ & Number of RISs and IUs respectively & $T_s$ & Slot duration\\
    \hline
     $\mathcal{K}$ & Number of elements in $i$-th RIS & $R_i$ & $i$-th RIS \\
     \hline
      $K_g$ & Number of elements in a subgroup & $N$ & Number of sources\\
      \hline
      $B$ & Number of subgroups of a RIS & $S_n(D_n)$ & $n$-th source (destination)\\
    \hline
    $T_{d_n}$ & Maximum acceptable delay for $S_n-D_n$ pair & $\sigma^2_0$ & Variance of the AWGN\\
      
    \hline
   $p_j$ & Priority of the $j$-th source & $\lambda_{u,k}(\mu_{u,k})$ & ON(OFF) period length of $k$-th IU \\
    \hline
    $\rho_L$ & Path-loss at one meter distance & $\alpha$ & Path-loss exponent\\
    \hline
    $d_{m,n}$ & Distance between nodes $m$ and $n$ & $h,g$ & Complex channel gain \\

    \hline
	$R_{r,b}$ & $b$-th subgroup of $r$-th RIS & $U_i$ & $i$-th IU\\
    \hline
    $P$, $P_{\rm phase},$ and $P_{\rm proc}$ & Transmit , phase shift, and processing power respectively & $T_c$ & Channel coherent time\\
    \hline
	$l$ & Euclidean distance from $S$ to $D$ & $K_b$ & Number of coded bits\\   
    \hline
    $W_{k,i}$ & Waiting time of $k$-th source at $U_i$ & $m_r$ & Modulation scheme \\
    \hline
    $\beta$ & A function of delay constraint and, channel condition & $\phi_{r,b}$ & phase shift for $R_{r,b}$\\ 
    \hline
    $T_{k,i}$ & Information transfer time for $k$-th $S-D$ pair at $i$-th IU  & $\alpha_k$ & Number of packets \\
    \hline
    $\tau_{k,i}$ & Processing time of $k$-th source at $i$-th IU & $\lambda$ & Wave length\\
    \hline
    $\eta_{w,i}$ & Waiting time (in slots) at a busy $i$-th IU  & $P_b$ & Bit error rate\\
    \hline
    $N_\kappa$ & Number of sources which satisfied $T_{d_{n,i}} \leq \kappa_{I,i}$ & $\mu_k$ & Service rate \\
    \hline
    $T_{d_{n,i}}$ &  Maximum waiting time at $i$-th IU of $n$-th source & $\gamma^{r,b}_{j,k}$ & signal-to-noise ratio\\
    \hline
    $\mu_{1,n}$  & Spatial correlation between $1$st and $n$-th element of $R_{r,b}$  & $R^{r,b}_{j,k}$ & Achievable data rate \\
    \hline
    $\kappa_{I,i}$ & Estimated idle time of $i$-th IU & ${\rm D_T}$ & Data throughput\\
    \hline
     $K_{p,q}$ & Rician factor of channel between $p$ and 
$q$ & $\delta$ & Phase shift error\\
    \hline
     $\rm E_C \;(E_{\rm eff})$ & Energy consumption (Energy efficiency) & $r$ & Coverage area\\
    \hline
  \end{tabular}
  }
  \vspace{-2mm}
\end{table*}

Each RIS has $\mathcal{K}$ reflecting elements, which are effectively controlled to adjust both the amplitude and phase of the incident waveform\footnote{Here we do not consider secondary reflections from the RISs. However, the proposed framework is general and it can be readily extended to the case where the aspect of secondary reflection is incorporated \cite{2hop}.}. However, for the sake of simplicity and mathematical tractability, the amplitude factor is set to unity and it is only the phase that is tuned or optimized \cite{partition3}. To reduce the channel estimation overhead, we employ a grouping strategy, where nearby $K_g$ reflecting elements are grouped together, such that all the elements belonging to the same group introduce identical phase shift \cite{partition3}. Specifically, the $r$-th RIS $R_r$, consisting of $\mathcal{K}$ reflecting elements, is subdivided into $B$ non-overlapping surfaces $R_{r,b}$ $\forall$ $1 \leq b \leq B$ such that $BK_g=\mathcal{K}$, where the number of partitions and the resulting sub-surfaces are determined a-priori \cite{partition3}. Without any loss of generality, we consider that each group has an equal number of elements and a particular group can serve one service request at a time \cite{can_serv_single_user}. In this scenario, $R_{r,b}$ has two possible states: ON and OFF. If an element is in ON state, the phase of an incident signal can be changed to a desired direction and it cannot reflect while in OFF state. Finally, an example of the considered scenario is demonstrated in Fig. \ref{sys}, where $S_1$ has the highest priority and the elements of the RIS are grouped into four non-overlapping surfaces, and the signal from $S_1$ reaches $D_1$ by using the path $S_1-U_1-R_{1,1}-U_3-U_4-D_1$.

\subsection{Channel Model}
Based on the IU availability, the $S_i-D_i$ pair communicates with or without the help of RISs.  We assume that the wireless links suffer from both large-scale path-loss effects and small-scale block fading. All the channels $S_i \rightarrow U_j$, $U_j \rightarrow U_k$, and $U_k \rightarrow D_i$  $\forall$ $j,k=1,\cdots,M$ experience Rician fading and their corresponding path-loss factors are $\rho_{\rm L}^{1/2}d_{S,U_j}^{-\alpha/2}$, $\rho_{\rm L}^{1/2}d_{U_j,U_k}^{-\alpha/2}$, and $\rho_{\rm L}^{1/2}d_{U_k,D}^{-\alpha/2}$, respectively, where $\rho_{\rm L}$ is the pathloss at one meter distance, $\alpha$ is the path-loss exponent and $d_{m,n}$ denotes the distance between $m$ and $n$. Specifically, we have the channel gain $h_{p,q}$ $\forall$ $\{p,q\}\in\{S_i,U_j,U_k,D_i\}$, where the distribution of $|h_{p,q}|$ is defined as\cite{tvt}
\vspace{-2mm}
\begin{align} \label{rice}
f_{|h_{p,q}|}(\alpha,K_{p,q})&=2(1+K_{p,q})e^{-K_{p,q}} \nonumber\\
& \!\!\!\!\!\!\!\!\!\!\!\!\!\!\!\!\!\!\!\!\!\!\!\!\!\!\!\!\!\!\!\! \times \alpha e^{-(1+K_{p,q})\alpha^2}I_0\left[ 2\alpha\sqrt{K_{p,q}(1+K_{p,q})}\right] , \:\: \alpha\geq 0.
\end{align}
Here $K_{p,q}$ is the Rician factor corresponding to the channel gain $h_{p,q}$ and $I_0(\cdot)$ denotes the zero-order modified Bessel function of the first kind. Moreover, note that direct $S-D$ path does not exist, i.e., $h_{S_i,D_i}=0$ $\forall$ $i$.

Since the elements of the RIS are divided into $B$ subgroups with $K_g$ reflecting elements on each of them, the composite channel between $S_i/U_j$ and $R_{r,b}$ is obtained as \cite{partition3}
\vspace{-2mm}
\begin{equation}  \label{hdef}
h_{S_i/U_j,R_{r,b}}=\sum\limits_{n \in R_{r,b}} h_n,
\end{equation}
where $h_n$ is the channel gain between $S_i/U_j$ and the $n$-th element of $R_{r,b}$ and $|h_n|$ follows identical distribution as \eqref{rice}. Similarly, the composite channel between $R_{r,b}$ and $U_k/D_i$ is
\vspace{-2mm}
\begin{equation}
g_{R_{r,b},U_k/D_i}=\sum\limits_{m \in R_{r,b}} g_m.
\end{equation}
Finally, as the reflecting elements within $R_{r,b}$ are very close to each other, it is practical to assume that the associated channels are spatially correlated. Note that in this work, we adopt the \emph{sinc} model \cite{corr} to mathematically characterize the spatial correlation at the RISs.

\subsection{User Traffic Characterization}  \label{utc}
In a typical wireless communication scenario, data generally arrives in bursts to the users. As a result, the IUs $U_1,\cdots,U_M$, in this work, are characterized by exponentially distributed OFF and ON period lengths, with means $\lambda_{u,k}$ and $\mu_{u,k}$, respectively. Without any loss of generality, $T_s$ is assumed to be small in comparison with $\mu_{u,k}$ and $\lambda_{u,k}$ \cite{crn}, which prevents $U_k$ $\forall$ $k=1,\cdots,M$ changing its status multiple times within a single $T_s$. Thus, the IU activities are characterized by a discrete-time Markov chain (DTMC) with the state transition probabilities \cite{crn}:
\vspace{-2mm}
\begin{equation}
\text{P}_{\text{ON}\rightarrow \text{OFF}}\overset{\Delta}{=}\text{p}_{10}=\int_0^{T_s}\frac{1}{\mu_{u,k}}e^{-a/\mu_{u,k}}da=1-e^{-T_S/\mu_{u,k}},
\end{equation}
\vspace{-2mm}
\begin{equation*}
\text{P}_{\text{OFF}\rightarrow \text{ON}}\overset{\Delta}{=}\text{p}_{01}=\int_0^{T_s}\frac{1}{\lambda_{u,k}}e^{-b/\lambda_{u,k}}db=1-e^{-T_S/\lambda_{u,k}}.
\end{equation*}
Accordingly the state transition matrix $\mathcal{P}$ is:
\begin{equation} \label{tpmtrx}
\mathcal{P}=
\begin{bmatrix}
   \text{p}_{00} & \text{p}_{01} \\
   \text{p}_{10} & \text{p}_{11} 
\end{bmatrix}=
 \begin{bmatrix}
   e^{-T_S/\lambda_{u,k}} & 1-e^{-T_S/\lambda_{u,k}} \\
   1-e^{-T_S/\mu_{u,k}} & e^{-T_S/\mu_{u,k}}
\end{bmatrix}.
\end{equation}

\section{Proposed strategy}\label{prop}
In this section, we propose a priority-aware scheduling strategy for grouping-based RIS-assisted wireless networks. Here, we consider a delay-constrained scenario, where the data from $S_i$ must arrive at $D_i$ within the delay bound $T_{d_i}$. Firstly, we characterize the achievable data rate for a grouping-based RIS-aided scenario. Secondly, based on the channel condition and the number of coding bits being used at the RIS, we comment on the suitable group selection strategy. Thirdly, depending on the channel condition as well as the delay sensitivity of the application, a novel scheduling strategy is proposed for information transfer between any two IUs. Note that, we also comment on the appropriate choice of modulation scheme and appropriate next hop selection criteria. Finally, we provide an illustrative example of the proposed framework.

\subsection{Achievable Rate Characterization of Grouping-based RIS}

Based on \eqref{hdef}, the adopted \emph{sinc} correlation model, and the theory of correlated random variables \cite{cmodel}, we obtain
\vspace{-2mm}
\begin{align}  \label{chandef}
h_1&=h_{I1}+jh_{Q1} \nonumber \\
\vdots& \nonumber \\
h_n&=\left(\mu_{1,n}h_{I1}+\sqrt{1-\mu_{1,n}^2}h_{In}\right) \nonumber \\
&\:\:+j\left(\mu_{1,n}h_{Q1}+\sqrt{1-\mu_{1,n}^2}h_{Qn}\right) \quad \forall \quad n \in R_{r,b},
\end{align}
where $\mu_{1,n}=\dfrac{\sin \left( \frac{2\pi}{\lambda}d_{1,n} \right)}{\frac{2\pi}{\lambda}d_{1,n}}$ is the spatial correlation and $d_{1,n}$ is the Euclidean distance between the first and $n$-th element of $R_{r,b}$ \cite{corr}. Moreover, $h_{In},h_{Qn}$ are independent Gaussian random variables denoting the in-phase and quadrature components of $h_n$, such that $|h_n|$ follows identical distribution as \eqref{rice}. Accordingly, from \eqref{hdef}, we can derive the composite $S_i/U_j-R_{r,b}$ channel  $h_{S_i/U_j,R_{r,b}}$ as
\vspace{-2mm}
\begin{align}  \label{hsum}
h_{S_i/U_j,R_{r,b}}&= \left(h_{I1}\sum\limits_{p=1}^{K_g} \mu_{1,p} + \sum\limits_{p=1}^{K_g} h_{Ip}\sqrt{1-\mu_{1,p}^2}\right) \nonumber \\
& + j \left( h_{Q1}\sum\limits_{p=1}^{K_g} \mu_{1,p} + \sum\limits_{p=1}^{K_g} h_{Qp}\sqrt{1-\mu_{1,p}^2} \right)
\end{align}
and similarly, the composite $R_{r,b}-U_k/D_i$ channel $g_{R_{r,b},U_k/D_i}$ can also be obtained\footnote{Statistical characterization of both the composite channels is straightforward as it only involves the aspect of the sum of correlated random variables.}. Hence, if a communication link exists between $U_j$ and $U_k$ via $R_{r,b}$, the resulting end-to-end achievable rate is expressed as
\vspace{-2mm}
\begin{equation}  \label{rate}
    R_{j,k}^{r,b}=\left(1-\frac{2}{T_c}\right) \log_2 \left( 1+\gamma_{j,k}^{r,b} \right),
\end{equation}
where the signal-to-noise ratio (SNR) $\gamma_{j,k}^{r,b}$ is defined as
\vspace{-2mm}
\begin{align} \label{SNR}
    \gamma_{j,k}^{r,b}&=\frac{P \rho_{\rm L}^2 \left(d_{U_j,R_r}d_{R_r,U_k}\right)^{-\alpha}}{\sigma_0^2} \Big| h_{U_j,R_{r,b}}\times g_{R_{r,b},U_k} \times e^{j\phi_{r,b}}  \Big|^2 \nonumber \\
    &= \zeta
    \Big| h_{U_j,R_{r,b}}\times g_{R_{r,b},U_k} \times e^{j\phi_{r,b}}  \Big|^2,
\end{align}
where $\zeta=\frac{P \rho_{\rm L}^2 \left(d_{U_j,R_r}d_{R_r,U_k}\right)^{-\alpha}}{\sigma_0^2}$, $\phi_{r,b}$ is the associated phase shift provided at $R_{r,b}$, $T_c$ is the channel coherence time normalized to the number of slots, and $\sigma_0$ is the power of the zero mean additive white Gaussian noise (AWGN) at $U_k$. Moreover, the quantity $\frac{2}{T_c}$ denotes the feedback and processing delay. Note that unlike the conventional RIS-based approach, $\gamma_{j,k}^{r,b}$ does not involve a diagonal phase shift matrix of $K_g$ non-zero elements. On the contrary, all the elements of $R_{r,b}$ provide a common phase shift $\phi_{r,b}$ to the incoming signal. Hence, by rewriting the complex composite channels as $h_{U_j,R_{r,b}} \!=\!\Big| h_{U_j,R_{r,b}}  \Big| e^{-j\theta_{h}}$ and $g_{R_{r,b},U_k} \!=\!\Big| g_{R_{r,b},U_k}  \Big| e^{-j\theta_{g}}$, we obtain the optimal phase shift corresponding to $R_{r,b}$ as $\phi_{r,b}^{\rm opt}=\theta_h+\theta_g$ such that the maximum attainable SNR is given by \eqref{optsnr}. Finally, as we focus on the performance of idle IUs acting as DF relays,  we consider an ideal scenario where we ignore the interference from other IUs; interference mitigation can be achieved through sophisticated signal processing and equalization techniques \cite{intrfrnc}.
\begin{figure*}
\begin{align}  \label{optsnr}
    \gamma_{j,k}^{r,b}&=\frac{P \rho_{\rm L}^2 \left(d_{U_,R_r}d_{R_r,U_k}\right)^{-\alpha}}{\sigma_0^2} \!\Big| h_{U_j,R_{r,b}}  \Big|^2\Big| g_{R_{r,b},U_k} \Big|^2 \nonumber \\
    &\overset{(a)}{=}\frac{P \rho_{\rm L}^2 \left(d_{U_j,R_r}d_{R_r,U_k}\right)^{-\alpha}}{\sigma_0^2} \left( \left(h_{I1}\sum\limits_{p=1}^{K_g} \mu_{1,p} + \sum\limits_{p=1}^{K_g} h_{Ip}\sqrt{1-\mu_{1,p}^2}\right)^2 +\left( h_{Q1}\sum\limits_{p=1}^{K_g} \mu_{1,p} + \sum\limits_{p=1}^{K_g} h_{Qp}\sqrt{1-\mu_{1,p}^2} \right)^2\right) \nonumber \\
    & \times \left( \left(g_{I1}\sum\limits_{p=1}^{K_g} \mu_{1,p} + \sum\limits_{p=1}^{K_g} g_{Ii}\sqrt{1-\mu_{1,p}^2}\right)^2 +\left( g_{Q1}\sum\limits_{p=1}^{K_g} \mu_{1,p} + \sum\limits_{p=1}^{K_g} g_{Qi}\sqrt{1-\mu_{1,p}^2} \right)^2\right),
\end{align}
where $(a)$ follows from \eqref{hsum}.

\vspace{-2mm}
\hrulefill
\vspace{-2mm}
\end{figure*}
\vspace{-2mm}
\begin{rem}
We observe from \eqref{rate} and \eqref{optsnr} that the achievable data rate $R_{j,k}^{r,b}$ is dependent on both the number of reflecting elements $K_g$ and the spatial correlation $\mu_{1,n}$ $\forall$ $n=1,\cdots,K_g$. Specifically,  $\mu_{1,n}$ is dependent on the Euclidean distance between the first and $n$-th element of the non-overlapping surface $R_{r,b}$ of the RIS $R_r$.
\end{rem}

\subsection{Group Selection Criteria at an RIS}  \label{discrete}

As the RIS is partitioned into $B$ non-overlapping subsurfaces, the question of appropriate selection arises in the context of multiple subgroups availability. Hence, here we discuss the group selection criterion at a particular RIS. In case a communication link exists between $U_j$ and $U_k$ via $R_r$,  the resulting rate, when the $b$-th group of $R_r$ is opted for, is given by \eqref{rate}.  By rewriting \eqref{SNR}, we obtain
\vspace{-2mm}
\begin{align}
    \gamma_{j,k}^{r,b}&=\zeta
    \Big| h_{U_j,R_{r,b}}\times g_{R_{r,b},U_k} \times e^{j\phi_{r,b}}  \Big|^2,
\end{align}
where from \eqref{hsum}, we have
\vspace{-2mm}
\begin{align}
\!\! h_{U_j,R_{r,b}} \!\!&=\Bigg(h_{I_1}\sum_{p=1}^{K_g}\mu_{1,p}+h_{I_1}\sum_{p=1}^{K_g}\sqrt{1-\mu^2_{1,p}}\Bigg) \nonumber \\
&+j\Bigg(h_{Q_1}\sum_{p=1}^{K_g}\mu_{1,p}+h_{Q_1}\sum_{p=1}^{K_g}\sqrt{1-\mu^2_{1,p}}\Bigg) \nonumber \\
&=\Bigg\{\Bigg(h_{I_1}\sum_{p=1}^{K_g}\mu_{1,p}+h_{I_1}\sum_{p=1}^{K_g}\sqrt{1-\mu^2_{1,p}}\Bigg)^2
\nonumber \\
&+\Bigg(h_{Q_1}\sum_{p=1}^{K_g}\mu_{1,p}+h_{Q_1}\sum_{p=1}^{K_g}\sqrt{1-\mu^2_{1,p}}\Bigg)^2\Bigg\}^{\frac{1}{2}} \nonumber \\
&  \times \arctan  \!\!\left(\dfrac{h_{Q_1}\displaystyle\sum_{p=1}^{K_g}\mu_{1,p}+h_{Q_1}\sum_{p=1}^{K_g}\sqrt{1-\mu^2_{1,p}}}{h_{I_1}\displaystyle\sum_{p=1}^{K_g}\mu_{1,p}+h_{I_1}\sum_{p=1}^{K_g}\sqrt{1-\mu^2_{1,p}}}\right)\!\!.
\end{align}
Similarly, we can represent $g_{R_{r,b},U_k}$ and hence, the optimal phase shift for the desired signal is
\vspace{-2mm}
\begin{equation}
    \phi^{opt}_{r,b} \!= \!\!\!\! \sum_{p \in \{h,g\}} \!\!\!\! \arctan \!\! \left(\!\!\frac{p_{Q_1}\displaystyle\sum_{p=1}^{K_g}\mu_{1,p}+p_{Q_1}\sum_{p=1}^{K_g}\sqrt{1-\mu^2_{1,p}}}{p_{I_1}\displaystyle\sum_{p=1}^{K_g}\mu_{1,p}+p_{I_1}\displaystyle\sum_{p=1}^{K_g}\sqrt{1-\mu^2_{1,p}}}\!\!\right)
\end{equation}\label{optphase}
Based on the considered grouping framework, $\phi^{opt}_{r,b}$ is the optimal phase shift corresponding to all the elements of $R_{r,b}.$  Here, we state to differentiate our grouping framework from the one presented in \cite{aerial}. This work considers a RIS-assisted aerial-terrestrial communication system, where the RIS is attached to a building, assisting the downlink communications of multiple UAV-user pairs. Here, the RIS is sub-divided into multiple groups, such that, one RIS group can serve one UAV-user pair. However, within a group, each reflecting element is allotted its own unique phase shift. On the contrary, in our proposed framework, all the elements belonging to the same group introduce identical phase shift, resulting in reduced channel estimation overhead.

Note that, in all practical cases, the phase shift $\phi^{c}_{r,b}$ is not a continuous quantity but it is chosen from the finite set of predetermined phase shifts offered by the RIS \cite{how_much_phase}. Therefore, we choose $\phi^{c}_{r,b}$ among them, which is closest to the optimal $\phi^{opt}_{r,b}$, i.e., $\theta_h+\theta_g$. It may happen, that we have $\phi^{c}_{r,b} \neq \phi^{opt}_{r,b}$ and thus, we define the corresponding error $\delta$ as $\delta=\phi^{opt}_{r,b}-\phi^{c}_{r,b}$. By assuming that each subgroup of $R_b$ is $K_b$ bits coded, there are $2^{K_b}$ phase shifts available with a constant interval of $\dfrac{2\pi}{2^{K_b}}$, i.e., we have 
\vspace{-2mm}
\begin{equation}
    -\dfrac{2\pi}{2^{K_b+1}} \leq \delta<\dfrac{2\pi}{2^{K_b+1}}.
\end{equation}
Therefore, we obtain $\phi^{c}_{r,b}=\dfrac{\pi r}{2^{K_b-1}}$ where $\{r : r \in \mathbb{I} \:\: \rm{and} \:\: 0\leq r \leq 2^{K_b-1} \}$. As a result, the suboptimal, i.e., closest to the optimal, SNR $\hat{\gamma}_{j,k}^{r,b}$ is characterized as
\vspace{-2mm}
\begin{equation}\label{esnr}
     \hat{\gamma}_{j,k}^{r,b}=\zeta
    \Big| h_{U_j,R_{r,b}}\Big|^2 \Big|g_{R_{r,b},U_k}   \Big|^2 \Big|e^{-j\delta}\Big|^2
 \end{equation}
and accordingly, the corresponding achievable data rate is
\vspace{-2mm}
\begin{equation}  \label{rategrp}
    \hat{R}_{j,k}^{r,b}=\left(1-\frac{2}{T_c}\right)\log_2 \left( 1+\hat{\gamma}_{j,k}^{r,b} \right).
\end{equation}
Note that we have $\hat{R}_{j,k}^{r,b}=R_{j,k}^{r,b}$ when $\delta=0$. In case of multiple subgroup availability at $R_r$, we choose the $b^*$-th subgroup, where we obtain
\vspace{-2mm}
\begin{equation}
    b^*=\argmax_{b \in \mathbb{B}} \hat{R}_{j,k}^{r,b}\overset{(a)}{\equiv}\argmax_{b \in \mathbb{B}} \hat{\gamma}_{j,k}^{r,b}.
\end{equation}
Here $\mathbb{B}$ denotes the set of available subgroups at $R_r$ and $(a)$ follows from the monotonic nature of $\log_2(\cdot)$. It is worth observing that the selection of a particular subgroup does not necessarily imply $\delta=0$ for the same. The reason for this is attributed to the fact that, generally we have $\Big| h_{U_j,R_{r,b_1}}\Big| \neq \Big| h_{U_j,R_{r,b_2}}\Big|$ and $\Big|g_{R_{r,b_1},U_k}\Big|\neq \Big|g_{R_{r,b_2},U_k}\Big|$ for $b_1 \neq b_2$ and $b_1,b_2 \in \mathbb{B}$. Hence, the choice of the available subgroup with the minimum corresponding $\delta$ value does not always give the desired result.

\subsection{Scheduling Strategy at the IU}\label{idle1}

The current state of an IU plays a crucial role if we are to comment on whether its selection as a relay is appropriate for a particular source-destination pair. Hence, we consider two separate scenarios, where a particular IU is idle or busy and accordingly decide on the scheduling strategy.

\subsection*{C-I. IU is presently idle}

Here, we assume that none of the IUs are aware of the entire topology as they can only communicate within a circular range of radius $r$. As a result, by using the beacon transmission technique \cite{beacon1}, the idle IUs $U_i$ are first identified in that range. Thereafter, based on its traffic characteristics, we estimate the time $\kappa_{I,i}$ for which it will continue to remain idle. Accordingly, $\kappa_{I,i}$ is estimated as \cite{dramp}
\vspace{-2mm}
\begin{equation}\label{idle}
    \kappa_{I,i} = \lambda_i \ln \left( \frac{1}{1-\delta} \right),
\end{equation}
where $\delta$ is the acceptable error threshold and $\lambda_i$ depends on the traffic characteristics of $U_i$ (defined in Section \ref{utc}). During this time interval, $U_i$ acts as a relay for the other active IUs. Suppose, at any arbitrary point of time, there are $N_A (\leq N)$ sources $S_1,\cdots,S_{N_A}$ wanting to communicate with their respective destinations $D_1,\cdots, D_{N_A}$ via the IUs. The minimum number of hops required to send packets from $S_n$ to $D_n$ within a delay constraint $T_{d_n}$ is $\Psi=\displaystyle\biggl \lceil \frac{l}{r} \biggr \rceil$, where $l$ is the Euclidean distance from $S_n$ to $D_n$. Moreover, $U_i$ agrees to act as a relay only for those IU pairs, whose maximum possible delay at $U_i$, i.e., $T_{d_{n, i}}$ $\forall$ $n=1,\cdots, N_A$  is less than $\kappa_{I,i}$ and the remaining are discarded automatically. But, an idle IU can serve a single request at a time. Therefore, if there are $N_{\kappa}$ $(\leq N_A)$ candidates requesting for $U_i$, these sources are prioritized depending on their respective acceptable delay limits $T_{d_{n,i}}$ $\forall$ $n=1,\cdots,N_{\kappa}$. 

Accordingly, let $p_j$ $\forall$ $j=1,\cdots,N_{\kappa}$ be the priority of the $j$-th requesting pair. Also, we define a quantity $q_j$, which is associated with the channel condition while transmitting a signal from $U_{i}$ to the next relay node of the $S_j-D_j$ pair. Accordingly, to decide among the candidate requesting pairs at $U_i$, we define a quantity $\beta$ as a convex combination of priority and the channel condition, i.e.,
\vspace{-2mm}
\begin{equation}  \label{cdef}
    \beta=cp_j+(1-c)q_j \qquad {\rm where} \qquad c \in[0,1].
\end{equation}
The choice of $c$ depends on the application under consideration. Finally, the candidate with maximum $\beta$ is selected and by continuing this process, the transmitted packet reaches its destination within a delay bound. To obtain more analytical insights, we consider the above-mentioned extreme scenarios, i.e., $c=0,1$.

\subsubsection{When $c=0$}
This denotes a channel-aware delay unconstrained scenario, where $U_i$ always chooses the candidate with best channel condition, irrespective of its priority status. In case the connection does not involve any RIS, adaptive modulation (discussed in Section \ref{adp}) is employed and the corresponding data rate is determined. Note that, the scenario of considering $c=0$ and not involving any RIS, essentially represents the classical problem of channel-aware scheduling in a multiple-user scenario \cite{iccmac}.

\subsubsection{When $c=1$}
This implies that unaware of the channel conditions, the selection decision is solely based on the priority of the requesting candidates. 

In this case, suppose $S_j-D_j$ is the selected source-destination pair with $T_{d_j}$ being the associated maximum acceptable delay. The computation of the corresponding data rate is discussed in Section \ref{ufwork}.

\subsection*{C-II. $U_i$ is currently busy but the time after which it becomes idle can be estimated}

There may be instances where an idle IU is unavailable presently. If $U_i$ becomes available after a certain time, it may be able to assist in signal transfer for a requesting pair. Therefore, based on the user traffic characteristic, we estimate the time $\eta_{w,i}$ after which $U_i$ is expected to become idle, i.e., after $\eta_{w,i}$ time slot, $U_i$ will be available. Consequently, $\eta_{w,i}$ is estimated as \cite{dramp}
\vspace{-2mm}
\begin{equation} \label{iwait}
    \eta_{w,i} = {\mu_i} \ln \left( \left(\frac{1-e^{1/\mu_i}}{p_{\rm th}} \right)^+\right) +1.
\end{equation}
where $\mu_i$ depends on user traffic characteristic of $U_i$, $p_{\rm th}$ is an acceptable threshold probability, and $x^+=\max(x,1)$. As this is a cooperative framework, after $\eta_{w,i}$ time slot, $U_i$ becomes idle and it agrees to serve as a relay node. Hereafter, the quantity $\kappa_{I,i}$ from \eqref{idle} is estimated. However, in this context, $U_i$ agrees to act as relay only for those IU pairs, whose maximum possible delay at $U_i$ is less than $\eta_{w,i}+\kappa_{I,i}$. We assume that there are $N_{\eta\kappa}(\leq N_A)$ requesting pairs and the approach as to how a requesting pair is served by the $U_i$ is similar to what is described in the previous subsection.

\begin{rem}
Note that, in this work, the scheduling depends on the weighted combination of the priority and channel conditions. However, if one is eager to look into some other parameters, the extension is trivial. Specifically, if there are $n$ parameters of concern, we have
\vspace{-2mm}
    \begin{equation}
        \beta=\sum\limits_{i=1}^nc_ix_i,
    \end{equation}
where $\displaystyle\sum\limits_{i=1}^nc_i=1$ and $c_i \in [0,1]$ $\forall$ $i=1,\cdots,n$.
\end{rem}

\subsection{Adaptive Modulation at the Intermediate Users}\label{adp}
Based on $\beta$, a particular service request is chosen by an IU. Therefore, the IU forwards the requesting pair's data to the subsequent hop, and a strategic choice of the subsequent hop is essential in the delay constraint communication scenario. However, to select the subsequent hop, we take into account the achievable data rate at the next IU. We accomplish this by utilizing an IU-IU connection with an adaptive modulation scheme. Note that, adaptive modulation is employed if and only if a connection is established between consecutive IUs without the involvement of any RIS. Now, we go into more detail about the adaptive modulation methods below.

The primary objective, in this case, is not simply to transfer data from $S_i$ to $D_i$, but also to take care of the delay constraint. Hence, in this work, we consider channel adaptive communication \cite{goldsmith2005wireless} with fixed transmission power at the IUs. However, the IUs use rate adaptation if and only if they are in direct LoS with one another and no RIS is being used to link them. Let $m_r$ be the modulation scheme for an IU, where $m_r\in\mathbf{M}=\{m_1,m_2,\cdots,m_{|\mathbf{M}|}\}$ and corresponding data rate is $D_r=\log_2({m_r})\; \forall \;r=1,2,\cdots,|\mathbf{M}|$. The choice of $m_r$ depends on the wireless channel, where the complex channel gain $h$, bit error rate (BER) $P_b$, constellation size $m_r$, and the received power $P\rho_{\rm L}d^{-\alpha}|h|^2$ are related as \cite{goldsmith2005wireless}:
\vspace{-2mm}
\begin{equation} \label{mqam}
P_b=c_1 \exp \left( \frac{-c_2 P\rho_{\rm L}d^{-\alpha}|h|^2}{\sigma^2 (m_r^{c_3}-c_4)} \right).
\end{equation}
Here $\sigma^2$ is the noise power, and $c_1,\cdots,c_4$ are modulation-specific constants, respectively. The above equation explains the relation between the chosen modulation scheme and the application-specific acceptable BER. The channel configuration between two IUs determines $m_r$. On the chosen $m_r$, for a complete transfer of $\alpha$ packets with $\phi$ bits each, it take $\tau(r)= \Bigl\lceil\dfrac{\alpha\phi}{D_r}\Bigr\rceil$ slots, where $\Bigl\lceil\cdot\Bigr\rceil$ denotes the ceiling function.

\subsection{Next Hop Selection Criteria}
\begin{figure*}
    \centering
    \includegraphics[width=0.84\linewidth]{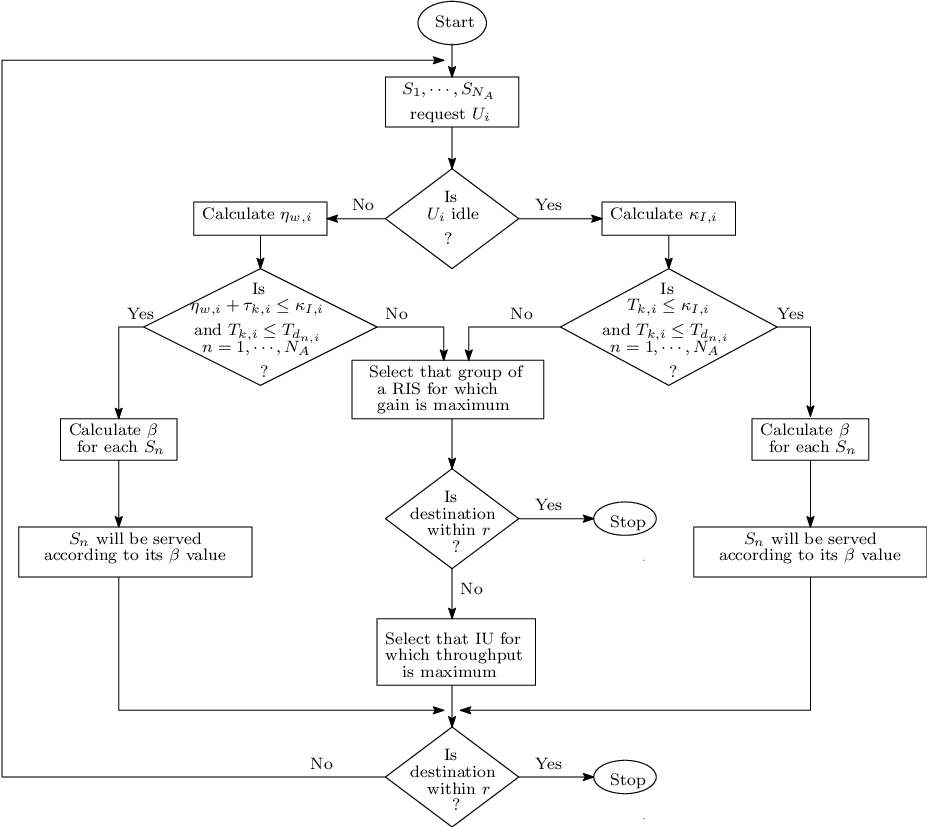}
    \vspace{-2mm}
    \caption{\footnotesize Flowchart of the proposed strategy.}
    \vspace{-4mm}
    \label{flowchart}
\end{figure*}

It is noted that, in the considered scenario of joint RIS and IU-assisted multihop framework, the classical LRD-based approach may not always yield the optimal routing solution. The LRD-based solution necessarily focuses on the fact that in the process of information transfer of an arbitrary $S-D$ pair, we always choose that particular IU/RIS at each hop, which minimizes the minimum distance towards $D$ from $S$. However, by doing so, we are overlooking aspects like the channel condition and/or the IU availability. As a result, at each hop, which can be an IU or RIS, only that candidate is chosen that has the best channel condition or sufficient idle time for which it can act as a DF relay or both, such that the respective delay constraint is not violated. In this context, an extensive delay analysis is provided in Section \ref{ufwork}. Depending on the channel condition and the availability of an IU, we go for RIS or IU. In our next hop selection process, by considering the channel condition and IU availability, we select that one as the next hop for which the achievable data rate is maximum. Accordingly, a particular algorithm is described in Fig. \ref{flowchart}. However, we demonstrate the justification of our claim later in the result section.

\subsection{Illustrative Example for Proposed Strategy}
The entire proposed strategy is presented in Fig. \ref{flowchart} in the form of a concise flowchart, which is illustrated below with the help of an example. As demonstrated in Fig. \ref{sys}, $N$ sources $S_1,\cdots,S_N$ want to communicate with their respective destination $D_1,\cdots,D_N$ via IU $U_1$. The sources are prioritized based on their $\beta$ values, i.e., acceptable delay limit and channel condition. We consider in this work that if $U_1$ is idle, it agrees to act as a relay, with the higher priority users being served first. In this figure, we assume $S_1$ to have the highest priority and hence, it is served prior to $S_2,\cdots,S_N$. Since there is no direct $U_1 \rightarrow U_2$ or $U_1 \rightarrow U_3$ LoS link, $U_1$ sends the packets to $U_2$ or $U_3$ via the RIS $R_{1,1}$. Here, we consider that the concerned RIS $R_1$ is already partitioned into four non-overlapping surfaces, $R_{1,1}, R_{1,2}, R_{1,3} $ and $R_{1,4}$, respectively. For communication purposes, $U_1$ selects $R_{1,1}$ to connect to $D_1$ via $U_2$ or $U_3$. It is interesting to note, that although LRD from $U_3$ to $D_1$ is greater than that from $U_2$ to $D_1$, $U_3$ is selected as the next hop. The reason for is attributed to the higher effective throughput in the case of $U_1 \rightarrow R_{1} \rightarrow U_3$ link as compared to $U_1\rightarrow R_1\rightarrow U_2$. Finally, $U_3$ forwards the information packets to $U_4$, which in turn forwards the same to $D_1$ and this completes the process.

\section{Analysis of the Proposed Scheduling Strategy} \label{ufwork}
As stated earlier in Section \ref{idle1}, here we investigate our proposed strategy in terms of delay-related metrics for information transfer from a source to a destination. The transmitted signals from a source reach their desired destination using IUs as well as RISs as hops. Now we go into more detail about the scenarios that are  discussed in Section \ref{idle1}.

When the $N_A$ sources place their request to $U_i$, depending on the considered value of $c$ in \eqref{cdef}, it calculates $\beta$ for all of them. Accordingly, the $\beta$s are arranged in decreasing order and maintained in a queue, with the source corresponding to the maximum $\beta$ getting service first. Therefore, the knowledge of $c$ is useful for  maintaining the queue of requested sources, i.e., which source will get service before or after the rest. With the arrival and the service rate of $U_i$ being denoted by $\lambda_{i,a}$ and $\mu_{i,i+1}$, respectively, the sources receive non-preemptive service in decreasing order of $\beta$. This implies that a particular source, if selected by $U_i$ for service, continues to receive the same, irrespective of the arrival of any other source (even with higher $\beta$). In case a source with higher $\beta$ arrives, it receives service only after the ongoing work has been completed. This results in a queuing delay, which is estimated below.

Here we first analyse the delay bound for a single hop and then characterize the same for complete information transfer between intended $S-D$ pair.
\subsection{$U_i$ Status Aware Single Hop Delay Bound}
Specifically, we consider two separate scenarios as follows.
\begin{enumerate}
    \item $U_i$ is presently idle
    \item $U_i$ is currently busy but the time after which it becomes idle can be estimated
\end{enumerate}

\subsection*{A-I: $U_i$ is presently idle} In this scenario, the source with maximum $\beta$ gets immediate service, i.e., no queuing delay. Suppose $W_{k,i}$ be the waiting time of the $k$-th requesting source $\forall$ $k=1,\cdots,N_{\kappa}$ at $U_i$. Hence, as $U_i$ is currently idle, for a source with highest $\beta$, we have $W_{1,i}=0$. For a source with the second largest $\beta$, $W_{2,i}$ depends on the processing time of the former. In general, if we denote the processing time of the $k$-th source at $U_i$ as $\tau_{k,i}(q_k)$, where $q_k$ depends on the modulation employed (discussed in Section \ref{adp}), we observe that $W_{k,i}$ is a function of $\tau_{2,i}(q_2),\cdots,\tau_{k-1,i}(q_{k-1})$, which in turn  depends on the channel condition of the corresponding source-destination pairs.
 Accordingly, the data transfer rate for the chosen pair is based on the discussion in Section \ref{adp}. Thereby, the processing time of $k$-th pair is
 \vspace{-2mm}
\begin{equation}  \label{trm1}
    \tau_{k,i}(q_k)=\frac{\alpha_k\phi_k}{\mu_{k}},
\end{equation}
where the $k$-th source needs to transfer $\alpha_k$ packets of data with $\phi_k$ bits in each of them and $\mu_k$ is the channel-dependent service rate. Hence, we obtain the corresponding $W_{k,i}$ as
\vspace{-2mm}
\begin{equation}  \label{trm2}
    W_{k,i}=\sum\limits_{j=1}^{k-1}\tau_{j,i}(q_j) =\sum\limits_{j=1}^{k-1} \frac{\alpha_j\phi_j}{\mu_{j}}
\end{equation}
Therefore, the total time required for the $k$-th pair at $U_i$ is the sum of $W_{k,i}$ and its own processing time, i.e., the total time for the information transfer is
\vspace{-2mm}
\begin{equation}  \label{trm3}
    T_{k,i}=W_{k,i}+\frac{\alpha_k\phi_k}{\mu_{k}}=\sum\limits_{j=1}^{k} \frac{\alpha_j\phi_j}{\mu_{j}}. 
\end{equation}
We assume that $U_i$ shares the information about its current status as well as the list of other sources waiting to be served, with each of the incoming requesting pair. Accordingly, the $k$-th incoming source waits for service at $U_i$ iff $T_{k,i} \leq \kappa_{I,i}.$ Note that the requesting sources are serviced at $U_i$ based on their corresponding $\beta$ values, where the processing time depends on the channel conditions. 

Hence, in the worst scenario, to guarantee a successful information transfer session, the entire process takes place with the minimum possible data rate, say binary phase shift keying (BPSK), with the service rate being $\mu_k=1$ $\forall$ $k$. Therefore, in this case, we obtain
\vspace{-2mm}
\begin{equation}
    \tau_{k,i}(q_k) =\alpha_k\phi_k \:\: {\rm and} \:\: W_{k,i}=\sum\limits_{j=1}^{k-1} \alpha_j\phi_j
\end{equation}    
from \eqref{trm1} and \eqref{trm2}, respectively. Hence, from \eqref{trm3} we obtain
\vspace{-2mm}
\begin{equation}  \label{trm4}
 T_{k,i}=\sum\limits_{j=1}^{k} \alpha_j\phi_j.
\end{equation}

The best case scenario implies that it is possible to process the data at the maximum possible data rate. If $\mu_{\max}$ be the corresponding service rate, we obtain
\vspace{-2mm}
\begin{equation} \label{mdata}
W_{k,i}=\frac{1}{\mu_{\max}}\sum\limits_{j=1}^{k-1}\alpha_j\phi_j\:\: {\rm and} \:\: T_{k,i}=\frac{1}{\mu_{\max}}\sum\limits_{j=1}^{k}\alpha_j\phi_j.
\end{equation} 
\begin{rem}  \label{rem3}
By considering the two extreme cases from \eqref{trm4} and \eqref{mdata}, we can state that
\vspace{-2mm}
\begin{equation}  \label{rem3eq}
\frac{1}{\mu_{\max}}\sum\limits_{j=1}^{k}\alpha_j\phi_j \leq T_{k,i} \leq \sum\limits_{j=1}^{k} \alpha_j\phi_j  \:\: \forall \:\: c \in [0,1].
\end{equation}
Note that the application-specific nature of the quantity $c$ is responsible in the calculation of $\beta$ at an IU. It does not have any impact on the bound on $T_{k,i}$ as stated above.
\end{rem}

\subsection*{A-II: $U_i$ is currently busy but the time after which it becomes idle can be estimated}
Here we first estimate the time interval $\eta_{w,i}$ from \eqref{iwait}, after which a particular IU becomes idle, given that it is currently busy. Note that, while the requesting pair with maximum $\beta$ was immediately getting service in the previous scenario, here it has to wait for time $\eta_{w,i}$, i.e., in this case, we have $W_{1,i}=\eta_{w,i}$ and not $W_{1,i}=0$ as previously discussed. Apart from this, the entire approach of estimating $W_{k,i}$ remains the same. Therefore, from \eqref{trm2} and \eqref{trm3}, we can aptly state that $\forall$ $c \in [0,1]$, we have
\vspace{-2mm}
\begin{align}  \label{etab}
    W_{k,i}&=\eta_{w,i}+\sum\limits_{j=1}^{k-1}\frac{\alpha_j\phi_j}{\mu_{j}} \:\: {\rm and} \:\: T_{k,i}=W_{k,i}+\frac{\alpha_k\phi_k}{\mu_{k}}.
\end{align}
As a result, similar to Remark \ref{rem3}, in this case we obtain
\vspace{-2mm}
\begin{equation}  \label{rem4eq}
    \eta_{w,i} +\frac{1}{\mu_{\max}}\sum\limits_{j=1}^{k}\alpha_j\phi_j \leq T_{k,i} \leq \eta_{w,i} +\sum\limits_{j=1}^{k} \alpha_j\phi_j  \:\: \forall \:\: c \in [0,1].
\end{equation}

Hence, we can summarize the complete subsection by generalizing the entire proposed framework as follows.
\vspace{-2mm}
\begin{equation}
W_{k,i} =
\begin{cases} 
a_i\eta_{w,i} + \displaystyle\frac{1}{\mu_{\max}}\sum\limits_{j=1}^{k-1}\alpha_j\phi_j & {\rm best \:\: scenario}, \\
a_i\eta_{w,i} + \displaystyle\sum\limits_{j=1}^{k-1} \alpha_j\phi_j & {\rm worst \:\: scenario}.
\end{cases}
\end{equation}
and
\vspace{-2mm}
\begin{equation}  \label{reseq2}
T_{k,i} =
\begin{cases}
a_i\eta_{w,i} + \displaystyle\frac{1}{\mu_{\max}}\sum\limits_{j=1}^{k}\alpha_j\phi_j & {\rm best \:\: scenario}, \\
a_i\eta_{w,i} + \displaystyle\sum\limits_{j=1}^{k} \alpha_j\phi_j & {\rm worst \:\: scenario}.
\end{cases}
\end{equation}
\noindent Here, $a_i=0$ implies that $U_i$ is idle and $a_i=1$ otherwise.

\subsection{Delay Bound for Complete Information Transfer for a Particular $S-D$ Pair}

Let for an arbitrary $S_m-D_m$ pair, the information from $S_m$ reaches $D_m$ via $n$ IUs $U_1,\cdots, U_n$. Hence, if the maximum acceptable waiting time and the actual waiting time at $U_i$ is $T_{d_{m, i}}$ and $t_{d_{m, i}}$, respectively, $\forall$ $i=0, \cdots, n$, we must have
\vspace{-2mm}
\begin{equation}  \label{appbnd}
    \sum_{i=0}^{n} T_{d_{m,i}} \leq T_{d_m},
\end{equation}
where $T_{d_m}$ is the maximum acceptable delay for the entire process of information transfer from $S_m$ to $D_m$. In the case of $t_{d_{m, i}} \leq T_{d_{m, i}}$ at $U_i$, the leftover waiting time $T_{d_{m, i}}-t_{d_{m, i}}$ is carried forward to $U_{i+1}$, i.e., the maximum acceptable waiting time $T_{d_{m,i+1}}$ at $U_{i+1}$ is now updated as $T_{d_{m,i+1}}+(T_{d_{m, i}}-t_{d_{m, i}})$ \cite{dramp}. This transfer of the `left-over' time at a particular hop to the consecutive ones is based on the fact that we consider a scenario, where the IUs do not have a global knowledge of the system topology, i.e., they cannot communicate beyond a distance $r$. Therefore, $U_i$ passes on $T_{d_{m, i}}-t_{d_{m, i}}$ to $U_{i+1}$, as it is unaware of exactly how many hops will be required for the complete information transfer.

Since the priority of $S_0-D_0$ pair at $U_i$ is inversely proportional to $T_{d_{0, i}}$, which in turn directly affects the calculation of $\beta$ value at $U_i$, we can rightly state that the priority parameter is not identical at each of the $n$ hops. For example, if the $S_0-D_0$ pair is the $k$-th candidate to be served at $U_i$ (based on the calculation of the $\beta$ values at $U_i$), this does not guarantee that it will also be the $k$-th candidate at $U_j$ $\forall$ $j \neq i$. We now provide a bound on the total time for information transfer corresponding to the $S_m-D_m$ pair, i.e., $\displaystyle\sum_{i=0}^{n} T_{d_{m,i}}$. From \eqref{etab}, we obtain the following.
\vspace{-2mm}
\begin{equation}  \label{delay_miu}
    \sum_{i=0}^{n} T_{d_{m,i}}= \sum_{i=0}^{n}a_i\eta_{w,i} + \sum_{i=0}^{n}\sum_{j=1}^{k_i} \frac{\alpha_{m}\phi_{m}}{\mu_{j}},
\end{equation}
where $a_i=0$ implies that $U_i$ is idle and $a_i=1$, otherwise. Here, $S_m$ intends to transfer $\alpha_m$ data packets of $\phi_m$ bits each to $D_m$ and this $S-D$ pair is served at the $k_i$-th position at the $i$-th hop with the corresponding service rate $\mu_{k_i}$. Note that, as discussed above, we do not guarantee $k_i \neq k_j$ $\forall$ $i \neq j$.
\begin{rem}  \label{rembnd}
If an arbitrary $S_m-D_m$ pair requires IUs $U_1,\cdots,U_n$ for complete information transfer, we have
\vspace{-2mm}
\begin{equation}
\frac{n\alpha_m\phi_m}{\mu_{\max}} \leq \sum_{i=0}^{n} T_{d_{m,i}} \leq \sum_{i=0}^{n}\eta_{w,i}+nk_{\max} \alpha_{m}\phi_{m},
\end{equation}
where $k_{\max}$ denotes the index of the last $S-D$ pair to be served by $U_i$ $i=1,\cdots,n$ such that \eqref{appbnd} holds. Note that the lower bound indicates the best case scenario, i.e., the $S_m-D_m$ pair finds all the IUs idle, gets served first among the competitors, and that also with the best channel condition. On the contrary, the upper bound implies that this $S-D$ pair has to wait at all the $n$ hops, is always the last to be served by the IUs, and that too with the worst possible channel condition.
\end{rem}
Finally, Remark \ref{rembnd} holds for the scenario, where IUs are available for each hop and the RISs in the surroundings are not being used at all. In cases where the RISs are being used due to IU unavailability, the only difference will be that the service rate $\mu_j$ in \eqref{delay_miu} will get replaced by $\hat{R}_{j,j+1}^{r,b}$ from \eqref{rategrp}. 

\section{Numerical Results}  \label{numerical}
In this section, we demonstrate the effectiveness of our proposed strategy and compare it with the existing benchmark schemes. Here we assume a Rician fading scenario, where the Rician factor is $K=10$ dB. To simulate the results, we consider the IU transmission power $P=30$ dBm, IU processing power $P_{\rm proc}=10$ dBm, phase shift power consumption, i.e, the power required to shift the phase of a RIS patch is $P_{\rm phase}=5$ dBm \cite{energyefficiency}, acceptable delay bound $T_d=50$ ms \cite{prmtr2}, path loss at one meter distance $\rho_L=10^{-3.53}$ \cite{prmtre1}, path loss exponent $\alpha=4.2$ between two consecutive IUs and $\alpha=2$ elsewhere, slot duration $T_s=100 \; \mu s$ \cite{crn} and wavelength $\lambda=0.1$ m \cite{grouping}. Moreover, a particular source $S$ sends $10^6$ data packets of unit length to its desired destination  $D$ and unless otherwise stated, we consider each RIS consisting of $400$ individual reflecting patches. Furthermore, we employ the M-ary quadrature amplitude modulation (M-QAM) technique with a BER of $P_b=10^{-6}$ as stated in Table \ref{tabrst}.
\begin{table} [!t]
\centering
  \caption{\footnotesize Transmission Modes for $P_b=10^{-6}$.} 
   \vspace{-2mm}
  \label{tabrst}
\resizebox{0.8\columnwidth}{!}{%
  \begin{tabular}{|c|c|c|}
    \hline \hline
    \textbf{SNR interval (dB)} & \textbf{Modulation} &  \textbf{Rate (bits/sym.)}\\
    \hline
     $(-\infty, 9.8554)$ & No transmission & 0 \\
    \hline
   $[9.8554, 12.8657)$ & BPSK & 1 \\
    \hline
    $[12.8657,14.6266)$ & QPSK & 2 \\
    \hline
    $[14.6266,15.8760)$ & $8$-QAM & 3 \\
    \hline
	$[15.8760,16.8451)$ & $16$-QAM & 4 \\
    \hline
    $[16.8451,17.6369)$ & $32$-QAM & 5 \\
    \hline
	$[17.6369,18.3063)$ & $64$-QAM & 6 \\   
    \hline
    $[18.3063, 18.8863)$ & $128$-QAM & 7 \\
    \hline
    $[18.8863, +\infty)$ & $256$-QAM & 8 \\
    \hline
  \end{tabular}
  }
  \vspace{-4mm}
\end{table}
Finally, we compare our strategy with the existing approaches.

\subsection{Impact of Grouping and Spatial Correlation}

\begin{figure*}[t]
 \begin{subfigure}[b]{.33\textwidth}
    \centering
    \includegraphics[width=0.92\linewidth]{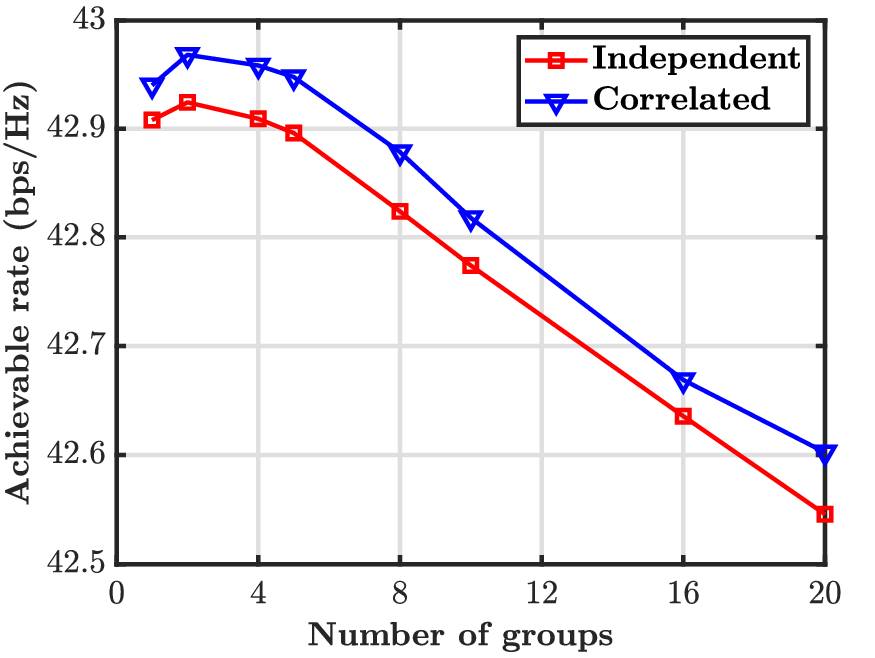}
    \vspace{-2mm}
    \caption{}
    \vspace{-2mm}
    \label{data rate vs no of groups}
\end{subfigure}
\begin{subfigure}[b]{.33\textwidth}
    \centering
    \includegraphics[width=0.92\linewidth]{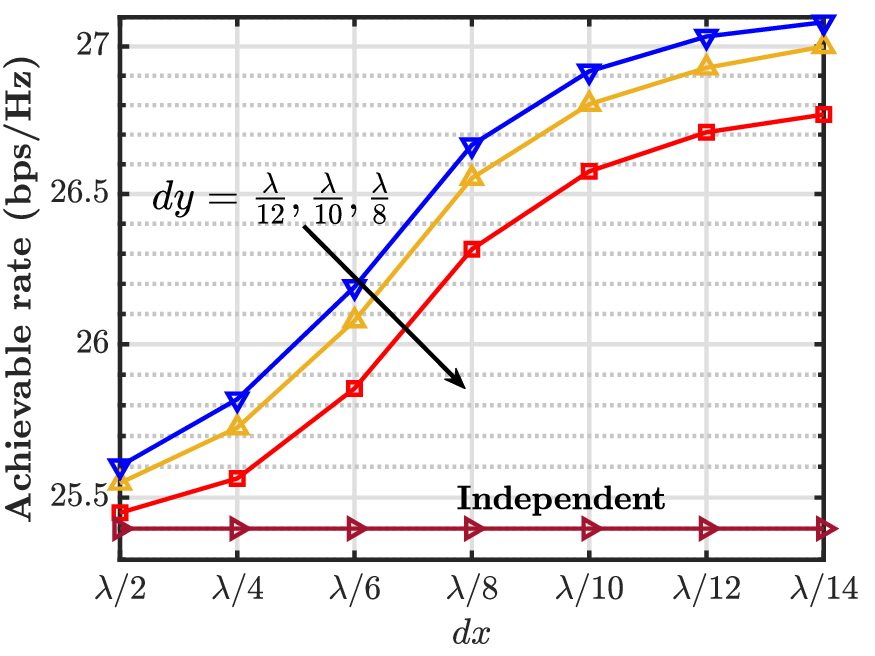}
    \vspace{-2mm}
    \caption{}
    \vspace{-2mm}
    \label{patch distance}
\end{subfigure}
\begin{subfigure}[b]{.33\textwidth}
    \centering
    \includegraphics[width=0.92\linewidth]{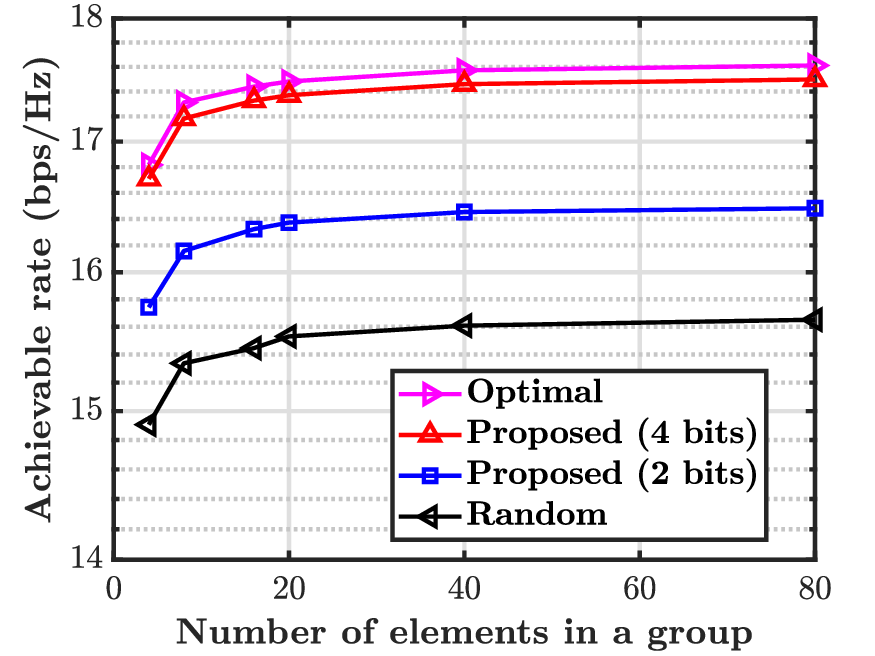}
    \vspace{-2mm}
     \caption{}
    \vspace{-2mm}
    \label{phase shift}
\end{subfigure}
\caption{\footnotesize  Impact of (a) Number of groups, (b) Patch spacing, and (c) Number of patches in a group on achievable data rate.}
\vspace{-6mm}
\end{figure*}

By accounting for the spatial correlation, here we demonstrate the advantage of grouping in terms of the achievable data rate. A particular RIS is sub-divided into groups with identical number of patches and patch spacing being $\lambda/8$ for the {\it correlated} scenario. Moreover, only one of these groups is used for the purpose of information transfer and the results corresponding to the {\it independent} channel scenario is obtained by setting the patch spacing to $\lambda/2$. Fig. \ref{data rate vs no of groups} illustrates the achievable data rate versus the number of groups into which the RIS has been sub-divided and this depicts the importance of incorporating the aspect of spatial correlation. Intuitively, this gain in performance can be understood from \eqref{hsum}, i.e., grouping essentially implies combination of channels corresponding to the adjacent patches. Hence, the combination is more likely to be constructive when the channels are correlated. We also observe, that the ideal group size is somewhere in between the two extreme scenarios of considering the group size to be unity or the total number of patches in the RIS. This further corroborates the trade-off investigated in \cite{partition3}  between the channel estimation overhead and the power gain offered by the group.

For a fixed number of patches in a group, Fig. \ref{patch distance} depicts the effect of the patch spacing on achievable data rate. In this context, we define the quantities $dx$ and $dy$, which denote the horizontal and vertical distance between consecutive patches, respectively. We observe that for a fixed $dx$, the achievable rate increases with decreasing $dy$ and vice-versa; for example, note the performance gap at $dx=\lambda/12$ between $dy=\lambda/12,\lambda/10,$ and $\lambda/8$. This justifies the effect of spatial correlation in \eqref{chandef}, which implies that the correlation is inversely proportional to the patch spacing. On the contrary, note that this aspect of patch distance does not have any impact in the \textit{independent} scenario, i.e., when  $dx=dy=\lambda/2$.

Fig. \ref{phase shift} shows the achievable data rate, defined in \eqref{rategrp}, as a function of the number of patches in a group. In this context, we consider groups with both {\it optimal} as well as {\it random} phases, where the optimal phase shift is calculated as in \eqref{optphase}. Moreover, as proposed in Section \ref{discrete}, we also take into account the scenario when the optimal phase shift is not a continuous quantity but it is chosen from a finite set of predetermined values. Specifically, in our {\it proposed} approach, we consider the group to be $2$ and $4$ bit coded, i.e., we choose the phase shift that is closest to the optimal from a set of $4(=2^2)$ and $16(=2^4)$ predetermined values, respectively. The figure demonstrates that, irrespective of the scenario, the achievable rate increases with the number of patches while the performance is worst in the case of random phases. Moreover, with an increase in the number of coded bits, the performance approaches the optimal. As can be seen, the achievable data rate with $4$ coded bits almost merges with its optimal counterpart.

\vspace{-2mm}

\subsection{Performance of Proposed Framework}
\begin{figure}[t]
    \centering
    \includegraphics[width=0.65\linewidth]{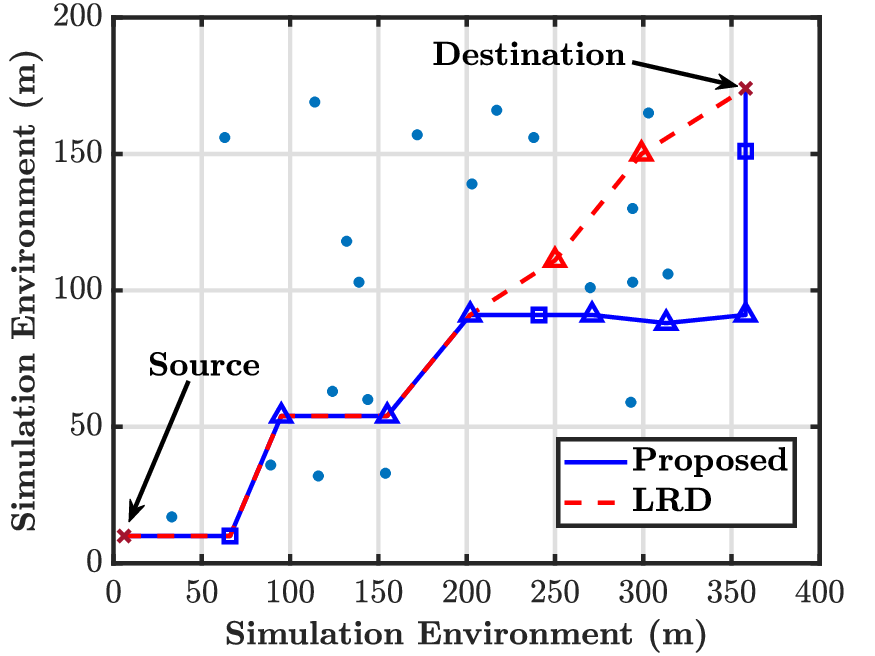}
    \vspace{-2mm}
    \caption{\footnotesize Trajectories for both the proposed framework and LRD-based approach; $\Box$ denotes RIS and $\Delta$ denotes IU.}
    \vspace{-6mm}
    \label{trajactory}
\end{figure}
An illustration of the proposed strategy based trajectory is presented in Fig. \ref{trajactory}. Here, we consider a two-dimensional squared area of $400 \times 400$ $\rm m^2$ where the RISs and IUs are placed strategically and randomly, respectively. Moreover, we assume that an IU has a coverage of $60$ meters, i,e., it cannot communicate beyond this distance. Furthermore, we consider a randomly selected $S-D$ pair in this grid, which is separated by a distance more than the coverage radius. Accordingly, a communication link is established both by the proposed strategy as well as the conventional LRD-based approach. The figure demonstrates that the proposed strategy requires more number of hops as compared to the other. However, in spite of this observation, to comment on the advantages of the former, we first define the following metrics.

\subsubsection{Data throughput $\rm D_T$ }
 We assume that the information transfer from $S$ to $D$ takes place by using $\aleph$ IUs and a certain number of RISs. Hence, $S$ transfers the data packets in $\aleph+1$ hops to $D$. Therefore, the data throughput is defined as
 \vspace{-2mm}
\begin{equation}\label{datath}
    {\rm D_T}=\dfrac{1}{\sum\limits_{i=1}^{\aleph+1} \dfrac{1-t_i}{\left( 1-P_b \right) m_{r_i}}+\dfrac{t_i}{R_i\left(\gamma_i\right)}},
\end{equation}
where
\vspace{-2mm}
\begin{align*}
t_i=\begin{cases} 
1, & \text{if $i$-th hop involves RIS} ,\\
0, & \text{else}.
\end{cases}
\end{align*}
Here $P_b$ is the BER, $m_{r_i}$ is the appropriate constellation size from Table \ref{tabrst}, and $R_i$ represents the achievable data rate from \eqref{rate}, when a RIS is chosen due to appropriate IU unavailability.

\begin{figure}
    \centering
    \includegraphics[width=0.68\linewidth]{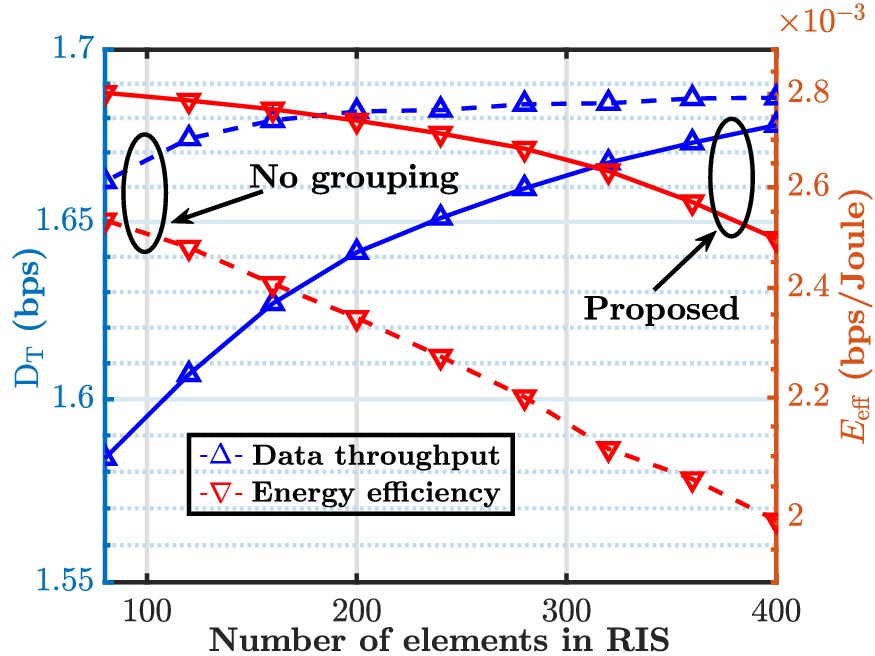}
    \vspace{-2mm}
    \caption{\footnotesize Performance trade-off investigation.}
    \vspace{-6mm}
    \label{nogrouping}
\end{figure}

\begin{figure*}[t]
 \begin{subfigure}{.33\textwidth}
    \centering
    \includegraphics[width=0.92\linewidth]{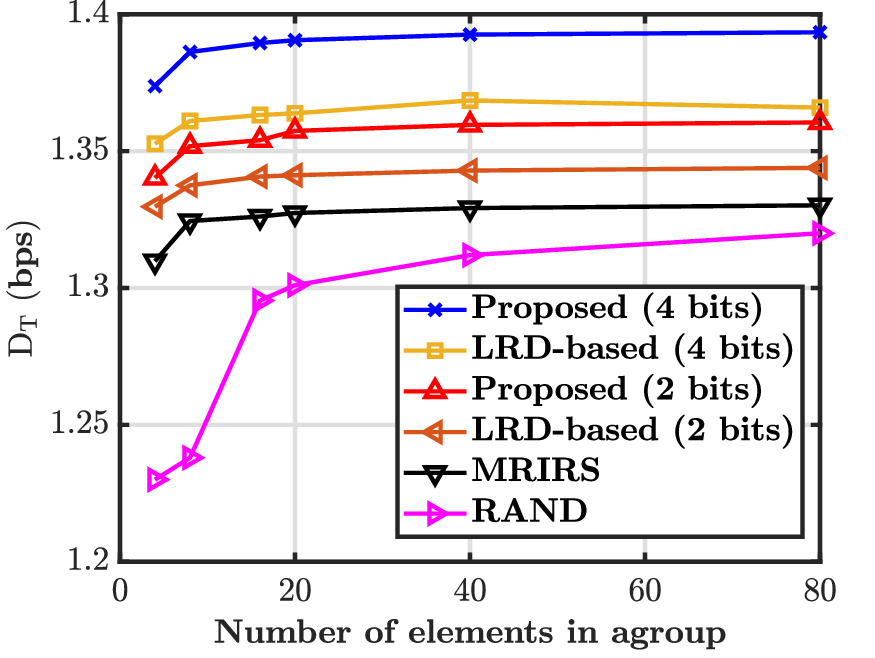}
    \vspace{-2mm}
    \caption{}
    \vspace{-3.2mm}
    \label{data throughput}
\end{subfigure}
\begin{subfigure}{.33\textwidth}
    \centering
    \includegraphics[width=0.92\linewidth]{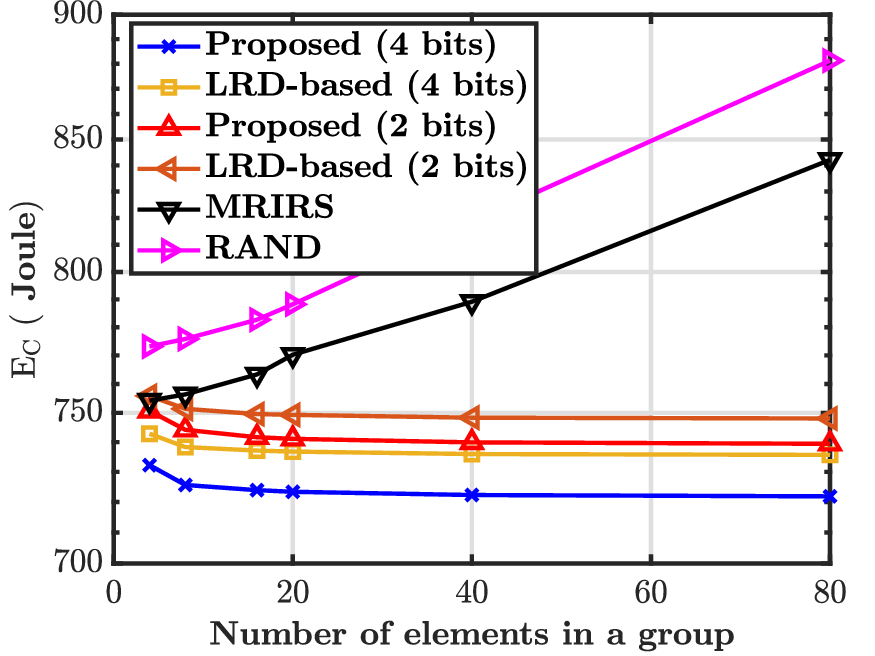}
    \vspace{-2mm}
    \caption{}
    \vspace{-3.2mm}
    \label{EC}
\end{subfigure}
\begin{subfigure}{.33\textwidth}
    \centering
    \includegraphics[width=0.92\linewidth]{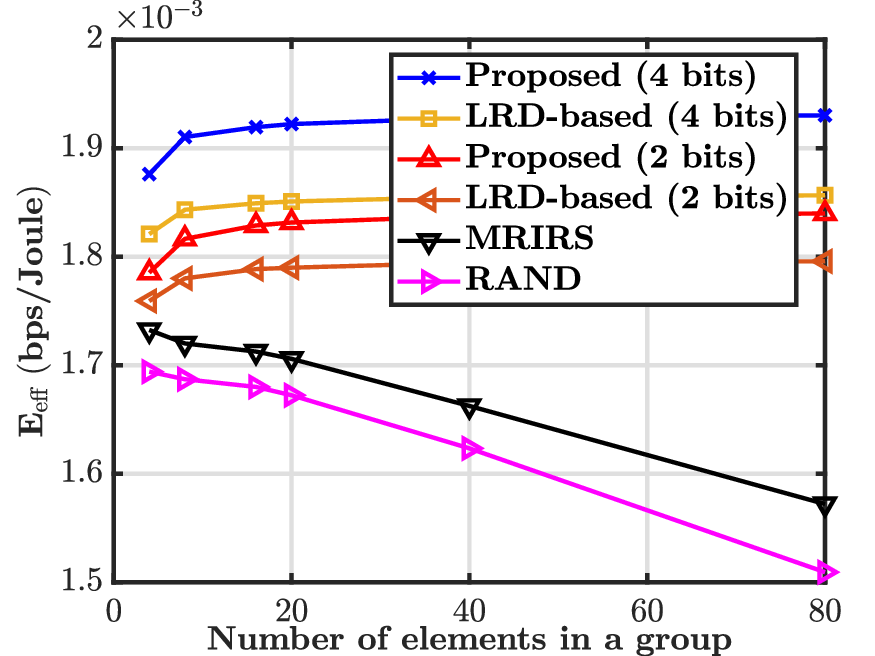}
    \vspace{-2mm}
    \caption{}
    \vspace{-3.2mm}
    \label{EE}
\end{subfigure}
\caption{\footnotesize Performance comparison: (a) Data throughput, (b) Energy consumption, (c) Energy efficiency.}
\vspace{-4mm}
\end{figure*}

\subsubsection{Energy Consumption ${\rm E_C}$}
The system energy consumption for sending $\alpha$ packets of data with $\varphi$ bits in each, where the $S-D$ pair connects in $\aleph+1$ hops, is defined as
\begin{align}\label{ec}
{\rm E_C}&= \alpha\varphi\sum\limits_{i=1}^{\aleph+1} \left(\dfrac{1-t_i}{\left( 1-P_b \right) m_{r_i}}+\dfrac{t_i}{R_i\left(\gamma_i\right)}\right)\nonumber\\
    &\times\left( P+s_iP_{\rm proc}+t_iP_{\rm phase}\right),
\end{align}
where $P_b,m_{r_i},t_i$ and $R_i\left(\gamma_i\right)$ are already defined above and
\vspace{-2mm}
\begin{align*}
s_i=\begin{cases} 
0, & \text{for} \; i=1,\\
1, & \text{else}.
\end{cases}
\end{align*}
Here, $P$, $P_{\rm proc}$, and $P_{\rm phase}$ are the transmission, processing and phase shift power consumption, respectively. Note that, $P_{\rm phase}$ is considered only if a particular hop involves a RIS.
\subsubsection{Energy Efficiency $E_{\rm eff}$}
Based on ${\rm D_T}$ and ${\rm E_C}$ from \eqref{datath} and \eqref{ec}, respectively, the corresponding trajectory energy efficiency is obtained as 
\vspace{-2mm}
\begin{equation}
     E_{\rm eff}=\frac{\rm D_T}{\rm E_C}.
\end{equation}
Based on these definitions, now we discuss the advantages of the proposed framework.

Fig. \ref{nogrouping} illustrates the importance of RIS grouping on the performance indicators, i.e., $\rm D_T$ and $\rm E_{eff}$, as a function of the number of elements in a RIS. Namely, we investigate the impact of considering a no grouping-based scenario (nGBS) and a grouping-based scenario (nGBS), where the RIS to be subdivided into four groups of equal size and the total number of patches in the RIS is varied accordingly. We observe that in both cases, $\rm D_T \;(\rm E_{eff})$ follows an increasing (decreasing) trend with the growing number of elements. This is because, growing number of patches supports better throughput. Note that, nGBS results in a higher data throughput as compared to GBS. This is intuitive as, unlike GBS, nGBS uses the entire RIS, which employs a fraction of the total number of patches in the RIS. Note that, nGBS limits the RIS usage to a single user, at any arbitrary point in time. On the contrary, GBS results in an enhanced energy efficiency performance. Hence, depending on the application at hand, for example, in a dense surrounding, we propose the usage of the GBS framework, which leads to better $\rm E_{eff}$ performance and also, more users can be catered to by the RISs at the same time.

Hereafter, we briefly discuss the existing benchmark schemes, which will be employed for the purpose of performance comparison.
\begin{enumerate}
\item LRD \cite{lrd2020}: A least remaining distance-based approach is considered, where factors such as wireless channel fluctuations are not taken into account. The sole objective here is to reduce the remaining distance to the destination in each hop.
\item MRIRS \cite{mrirs}: The work involves multihop route establishment between $S$ and $D$, without considering factors such as user traffic characteristics or channel aware adaptive modulation (AM).
\item RAND \cite{corr}: The work investigates the impact of random phase shifts and its impact on the corresponding achievable data rate. It doesn't entail any multihop communication between $S$ and $D$.
 \end{enumerate}

Fig. \ref{data throughput} depicts an overall increasing trend of $\rm D_T$ with number of RIS elements in a group, irrespective of the scheme considered. The reason behind this observation can be attributed to the fact that for hops involving a RIS, the achievable throughput is a function of the number of reflecting elements that are being used. As the proposed strategy is IU traffic aware, involves channel-dependent transmission techniques, and also considers the aspect of spatial correlation among the reflecting patches, it performs significantly better compared to its competitors. Moreover, the figure illustrates the fact that, as we are dealing with a delay-constrained scenario, an LRD-based approach is not always the best solution. It may result in a lesser number of hops (as observed in Fig. \ref{trajactory}) but at the cost of reduced data rate due to factors such as wireless channel fluctuations. Furthermore, here also we observe the impact of having more number of coded bits in the RIS.

Fig. \ref{EC} investigates the aspect of energy consumption among the proposed strategy and the existing benchmarks. It is interesting to note that ${\rm E_C}$, in the context of both our strategy and the LRD-based approach, decreases with an increasing number of elements in a RIS group. The reason behind
this counter-intuitive observation is that, in both these scenarios, irrespective of the group size, a single phase shift is required for the entire group and the LRD becomes a criteria only when the next hop choice is concerned. Moreover, as observed from \eqref{rate}, the achievable data rate increases with the group size. Hence, when we combine both these factors, it leads to this interesting insight, which can also be noted from \eqref{ec}. On the contrary, both for MRIRS and RAND, as individual phase shifts are required at all the patches, the energy consumption increases linearly with increasing group size.

Finally, Fig. \ref{EE} shows the impact of the proposed strategy on the system energy efficiency. We observe that, while ${\rm E_{eff}}$ for MRIRS and RAND monotonically decreases with the number of reflecting elements in a group, it follows an exact opposite trend in the context of our proposed strategy. The reason behind this observation is identical to the one as already described above in the discussion related to Fig. \ref{EC}.

\begin{figure}
    \centering
    \includegraphics[width=0.7\linewidth]{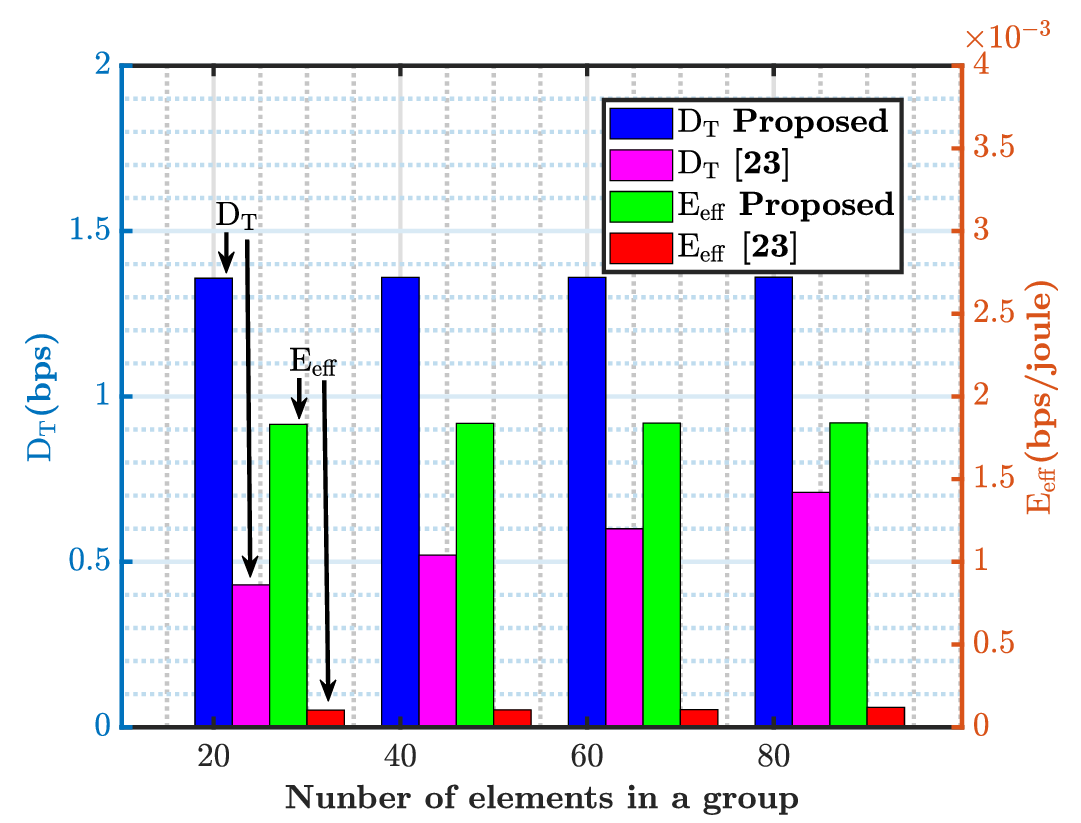}
    \vspace{-2mm}
    \caption{\footnotesize Impact of joint IU-RIS framework.}
    \vspace{-6mm}
    \label{ee_iotj}
\end{figure}

As discussed earlier, specifically the work in \cite{eecomp} investigates the aspect of energy efficiency maximization in the context of multihop RIS-assisted wireless networks. However, we argue, that there will be significant attenuation of the transmitted signal, if only RISs and no IUs are used to connect a transmitter to its intended receiver. Hence, we propose a novel joint IU-RIS framework, which also depends on the traffic characteristics of the IUs and the associated spatial correlation at the RIS. In this context, by varying the number of reflective elements in the group, Fig. \ref{ee_iotj} compares the $\rm D_T$ and $\rm E_{eff}$ performance of \cite{eecomp} and our proposed strategy. We observe here, that irrespective of the nature of variation of these performance metrics, our proposed strategy consistently and significantly outperforms \cite{eecomp}. This demonstrates the importance of incorporating the idle IUs to act as relays and not solely rely on the RISs deployed in the surroundings. Moreover, this also avoids unnecessary wastage of resources, i.e., dense RIS deployment.

\begin{figure*}[t]
 \begin{subfigure}[b]{.33\textwidth}
    \centering
    \includegraphics[width=0.92\linewidth]{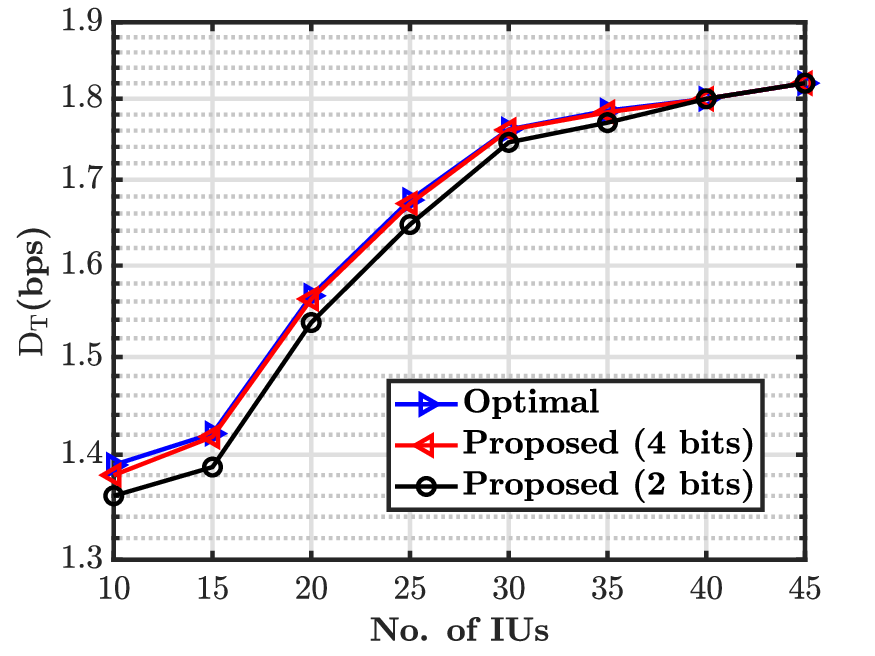}
    \vspace{-2mm}
    \caption{}
    \vspace{-2mm}
    \label{iudense}
\end{subfigure}
\begin{subfigure}[b]{.33\textwidth}
    \centering
    \includegraphics[width=0.92\linewidth]{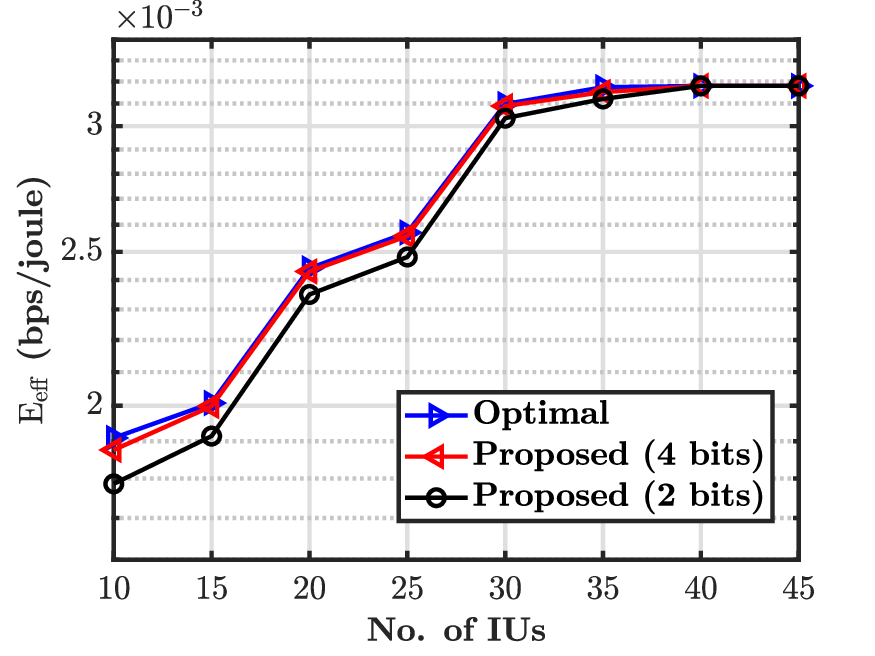}
    \vspace{-2mm}
    \caption{}
    \vspace{-2mm}
    \label{iudense_ee}
\end{subfigure}
\begin{subfigure}[b]{.33\textwidth}
    \centering
    \includegraphics[width=0.92\linewidth]{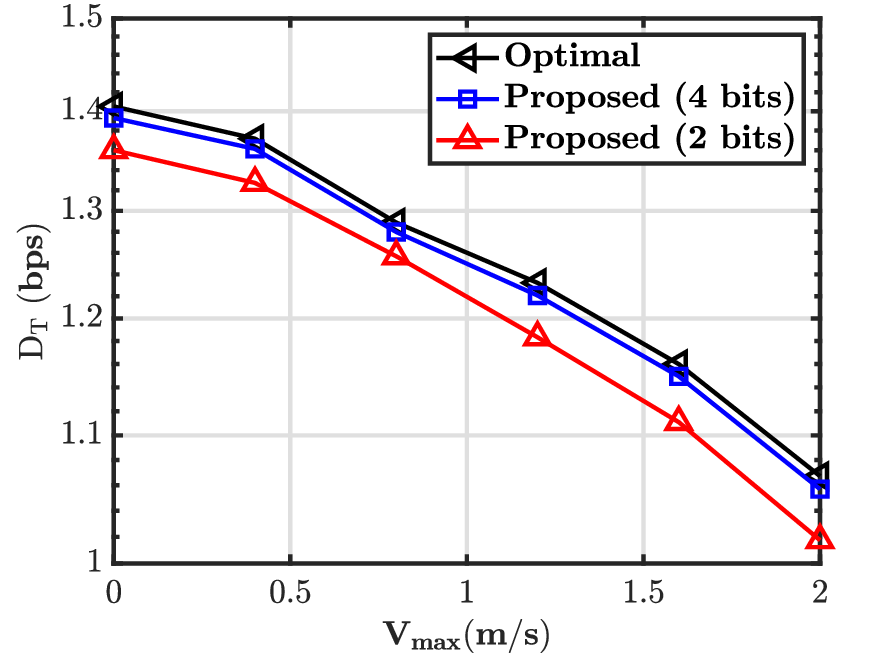}
    \vspace{-2mm}
     \caption{}
    \vspace{-2mm}
    \label{mob}
\end{subfigure}
\caption{\footnotesize  Impact of IUs: (a) No. of IUs vs. data throughput, (b) No. of IUs vs. energy efficiency, and (c) IU mobility vs. data throughput.}
\vspace{-6mm}
\end{figure*}


\subsection{Impact of IUs on System Performance}

In the proposed framework, Fig. \ref{iudense} investigates the aspect of data throughput as a function of the IU density. Here, we consider that each RIS consists of four identical groups with $\lambda/8$ spatial distance. We observe that, irrespective of the scheme, $\rm D_T$ increases with the growing number of IUs. This is also intuitive, as the growing number of IUs implies greater IU availability, which in turn, reduces the required number of hops for information transfer. As a result, fewer RISs are being used with increasing IU density. Moreover, it is interesting to note that, the number of RIS used in establishing the desired multihop connection asymptotically reaches zero with increasing number of IUs in the surroundings. This explains the saturating trend of $\rm D_T$ with the number of IUs. Furthermore, with no RISs being used with higher number of IUs, the corresponding $\rm D_T$ of all the considered schemes merge together, i.e., when RISs are not being used, it does not matter, whether RIS grouping is considered or not.

In an identical setup as in Fig. \ref{iudense}, Fig. \ref{iudense_ee} demonstrates the impact of the IU density on the system energy efficiency $\rm E_{eff}$. We observe that, for the scenarios as mentioned earlier, $\rm E_{eff}$ exhibits an increasing trend with the growing number of IUs. From \eqref{ec}, we note that, for a device pair to communicate via a RIS, an additional power consumption is incurred due to the required phase shift operation at the RIS. On the other hand, no such power consumption takes place if two devices communicate directly. As a result, a lesser amount of energy is required, i.e., a higher $\rm E_{eff}$. Similar to the previous figure, here too, the RIS dependency reduces with the increasing number of IUs, which also explains the saturating nature of the curve.

\subsection*{Impact of IU Mobility}

Here a mobile scenario is considered, i.e., the IUs have a particular velocity in a certain direction. It is to be noted that this will have a significant effect on the performance of the proposed scheme, as the inter-IU distance changes during the process of information transfer. In this context, Fig. \ref{mob} illustrates the impact of IU mobility on the proposed framework, with the coverage being taken as $60$ m and the mobility aspect being modeled by the standard random waypoint (RWP) model \cite{rwp}. Specifically, for this figure, we consider a pedestrian scenario \cite{mobility}, where we fix the maximum possible velocity $V_{\max}$ of the IUs, consider a random velocity chosen uniformly in $[0,V_{\max}]$ and observe its impact on the system data throughput. The figure illustrates that the performance deteriorates monotonically with increasing $V_{\max}$, which is intuitive. Such an observation is due to the mobility phenomenon, where a particular IU can move outside the coverage of another IU even during the communication process. This inevitably leads to outage, resulting in lesser data rate. Moreover, note that this figure with $V_{\max}=0$ is a special case corresponding to the static scenario, i.e., Fig. \ref{data throughput}. Furthermore, for a fixed $V_{\max}$, ${\rm D_T}$ decreases with the number of coded bits, which is also inline with the observation made in Fig. \ref{data throughput} as well.


\section{Conclusion}\label{con}
In this paper, we proposed a novel priority-aware channel-dependent scheduling strategy for RIS-assisted multihop D2D communication, which also takes into account the aspect of element grouping at the RIS. The proposed strategy avoids resource wastage by exploiting the very structure of the RIS, the associated spatial correlation, and also the randomness in the wireless channel. Moreover, we also claim that, in this context of RIS-aided multihop routing, the LRD-approach does not always yield the best result in terms of system performance. Finally, numerical results demonstrate the advantages of the proposed framework in terms of higher data rate, lower energy consumption, and higher energy efficiency with respect to the existing benchmark schemes. An immediate extension of this work is to investigate scenarios, where even when idle, it depends on the IUs to decide whether to act as relays to its requesting neighbors or to prioritize its own sleep mode. Moreover, we intend to investigate the aspect of imperfect channel state and phase errors at the RISs on the proposed strategy. Finally, we also aim to comment on having various grouping criteria at the RISs and investigate its impact.

\bibliographystyle{IEEEtran}
\bibliography{ref}

\end{document}